\documentclass[11pt,twoside]{article}
\usepackage{mathrsfs}
\usepackage{amsmath}
\usepackage{amssymb}
\usepackage{graphicx}
\usepackage{subfigure}
\usepackage[title]{appendix}
\usepackage{cite}
\numberwithin{equation}{section}

\newtheorem{theorem}{Theorem}
\newtheorem{proposition}{Proposition}
\newtheorem{lemma}{Lemma}

\newtheorem{corollary}{Corollary}
\newtheorem{remark}{Remark}

\DeclareMathOperator{\re}{Re}

\DeclareMathOperator{\diag}{diag}
\newcommand*{\QEDB}{\hfill\ensuremath{\square}}

\title{Integrable nonlocal finite-dimensional Hamiltonian systems related to the Ablowitz-Kaup-Newell-Segur system}
\author{
Baoqiang Xia and Ruguang Zhou
\\
School of Mathematics and Statistics, Jiangsu Normal University,
\\
Xuzhou, Jiangsu 221116, P. R. China,
\\
E-mail address: xiabaoqiang@126.com; zhouruguang@jsnu.edu.cn
}

\date{}
\begin{document}
\maketitle
\begin{abstract}

The method of nonlinearization of the Lax pair is developed for the Ablowitz-Kaup-Newell-Segur (AKNS) equation in the presence of space-inverse reductions. As a result, we obtain a new type of finite-dimensional Hamiltonian systems: they are nonlocal in the sense that the inverse of the space variable is involved. For such nonlocal Hamiltonian systems, we show that they preserve the Liouville integrability and they can be linearized on the Jacobi variety.
We also show how to construct the algebro-geometric solutions to the AKNS equation with space-inverse reductions by virtue of our nonlocal finite-dimensional Hamiltonian systems. As an application, algebro-geometric solutions to the AKNS equation with the Dirichlet and with the Neumann boundary conditions, and algebro-geometric solutions to the nonlocal nonlinear Schr\"{o}dinger (NLS) equation  are obtained.


\end{abstract}
\newpage

\section{ Introduction}
The technique of nonlinearization of the Lax pair (NLP) provides a powerful way to investigate soliton equations \cite{Cao88,Cao90,CaoGeng1990,CaoGengWu1999}.
By applying such a technique, a soliton equation in $(1+1)$-dimension is decomposed into two compatible finite-dimensional integrable Hamiltonian systems. As an application, some physically important solutions of soliton equations, such as quasi-periodic solutions, can be constructed through solving the corresponding finite-dimensional integrable Hamiltonian systems \cite{Zhou97,CGW1999,GDZ2007,Qiao1999,Qiao03}. A large amount of work and progress has been made in developing the technique of NLP; see e.g. \cite{Geng93,Zeng95,MaStr94,LiMa2000,Zhou2009,XuCaoNi21}.

However, the NLP method that allows to impose space-inverse reductions on the potentials is still an open problem to the best of our knowledge. Let us take the well-known Ablowitz-Kaup-Newell-Segur (AKNS) equation \cite{AKNS1974}
\begin{eqnarray}
\left\{\begin{array}{l}
u_t-iu_{xx}+2iu^2v=0,\\
v_t+iv_{xx}-2iuv^2=0,
\end{array}\right.
\label{akns}
\end{eqnarray}
as an example to illustrate this issue more clearly.
The NLP method has been successfully implemented to the AKNS equation (\ref{akns}) as early as in 1990 \cite{CaoGeng1990}.
The Liouville integrability of the resulting finite-dimensional Hamiltonian systems is ensured by confocal involution systems \cite{Moser1978,Moser1980}. However the known result in general does not allow to impose the following space-inverse reductions on the potentials:
\begin{eqnarray}
u(x,t)=\epsilon u(-x,t),~~v(x,t)=\epsilon v(-x,t),
\label{oddevenuv}
\\
u(x,t)=\epsilon\bar{v}(-x,t),
\label{nonlocalred}
\end{eqnarray}
where $\epsilon=\pm 1$ and the bar denotes complex conjugate.
We note that both of the above reductions have important applications. For example, once the odd reduction ((\ref{oddevenuv}) with $\epsilon=-1$) or even reduction ((\ref{oddevenuv}) with $\epsilon=1$) holds,
we obtain the potentials $u$ and $v$ that satisfy the Dirichlet boundary condition (see e.g. \cite{Sklyanin1987,Habibullin1991})
\begin{equation}
\left.u\right|_{x=0}=\left.v\right|_{x=0}=0,
\label{DBC}
\end{equation}
or the Neumann boundary condition
\begin{equation}
\left.u_x\right|_{x=0}=\left.v_x\right|_{x=0}=0,
\label{NBC}
\end{equation}
respectively, when the problem is restricted on the half-line.
Once the reduction (\ref{nonlocalred}) holds, we obtain a potential $u$ that solves the so-called nonlocal nonlinear Schr\"{o}dinger (NLS) equation introduced by Ablowitz and Musslimani in \cite{AM2013,AM2016}:
\begin{eqnarray}
u_t(x,t)-iu_{xx}(x,t)+2i\epsilon u^2(x,t)\bar{u}(-x,t)=0, ~~\epsilon=\pm 1.
\label{nnls}
\end{eqnarray}
Such a model was shown to be integrable (see e.g. \cite{AM2013,AM2016,Fokas2016,Caud2018,GS2017}) and has aroused enormous interest in recent years.

In this paper we develop further the method of NLP such that it permits to impose space-inverse reductions on the potentials. The main idea of our method is to appropriately choose eigenvalue parameters and eigenfunctions to realize the reductions in the potentials in the NLP process. By applying our method to the AKNS system in the presence of the aforementioned reductions, we obtain a new type of finite-dimensional Hamiltonian systems. The essential difference of our new finite-dimensional Hamiltonian systems, compared to the classical ones, is that they are nonlocal with respect to the space variable: they involve the inverse of the space variable (see section 3). We find that the Liouville integrability of our new finite-dimensional Hamiltonian systems survives. Particularly, we show that our nonlocal Hamiltonian systems can also be linearized on the Jacobi variety.
In this paper, we also study the applications of our nonlocal integrable finite-dimensional Hamiltonian systems to the constructions of solutions of the AKNS equation in the presence of space-inverse reductions.
As a consequence, we obtain algebraic-geometric solutions to the AKNS equation with Dirichlet/Neumann boundary conditions and obtain algebraic-geometric solutions to the nonlocal NLS equation (see section 4). The main technical difficulty in the construction of these solutions is the analysis of the symmetry relations involved in the parameters appearing in the solutions constructed by the standard NLP method. These symmetry relations result from the corresponding symmetries amongst the branch points of the hyperelliptic curve associated with our nonlocal Hamiltonian systems.

The paper is organized as follows. In section 2, we collect the main results on the standard NLP method to the AKNS equation that we will need. This also serves to fix notations. In section 3, we present a procedure of the NLP that permits to impose space-inverse reductions on the potentials. We show that the resulting nonlocal finite-dimensional Hamiltonian systems are Liouville integrable and they can be linearized by the Abel-Jacobi variables. In section 4, we discuss the applications of our nonlocal Hamiltonian systems to the constructions of algebro-geometric solutions to the AKNS equation with space-inverse reductions (\ref{oddevenuv}) and (\ref{nonlocalred}). We draw several concluding remarks in section 5. A few details on the derivations of some formulae are presented in appendices.

\section{Review of the NLP for the AKNS system}

In this section, we briefly review some essential results regarding the NLP method to the AKNS equation that we will need. We refer the reader to \cite{CaoGeng1990,CaoGengWu1999} for more details on this issue.

Let us first introduce the following notations
\begin{equation}
\mathbf{\Lambda}=diag(\alpha_1,\alpha_2,\cdots,\alpha_{2N}),~~\mathbf{p}=(p_{1},p_{2},\cdots p_{2N})^T,\quad \mathbf{q} =(q_{1},q_{2},\cdots q_{2N})^T,
\label{notationpq}
\end{equation}
where $\alpha_j$, $1\leq j\leq 2N$, are $2N$ distinct parameters.
We use the diamond bracket to stand for the standard Euclidean scalar product, for example
$$\langle \mathbf{p}, \mathbf{p} \rangle=\sum_{j=1}^{2N}p_j^2,~~\langle \mathbf{p},\mathbf{q} \rangle=\sum_{j=1}^{2N}p_jq_j.$$
We consider the standard symplectic space $(R^{4N},\omega)$ with the symplectic structure defined by
\begin{equation}
\omega=\sum_{j=1}^{2N}dp_j\wedge dq_j.
\label{symstruce}
\end{equation}
The Poisson bracket corresponding to this symplectic space is given by
\begin{equation}
\left\{f,g\right\}=\sum_{j=1}^{2N}\left(\frac{\partial f}{\partial
q_j} \frac{\partial g}{\partial p_j}-\frac{\partial f}{\partial p_j}\frac{\partial g}{\partial q_j}\right).
\label{BP}
\end{equation}

\subsection{Finite-dimensional integrable Hamiltonian systems related to the AKNS system}

The AKNS equation (\ref{akns}) is equivalent to the compatibility of the following Lax pair of linear spectral problems \cite{AKNS1974}
\begin{subequations}
\begin{eqnarray}
\phi_x=U\phi, ~~U=\left( \begin{array}{cc}
\frac{\lambda}{2} & u \\
 v &  -\frac{\lambda}{2} \\
 \end{array} \right),
 \label{lpx}
 \\
\phi_t=V\phi,~~V=i\left( \begin{array}{cc}
 \frac{\lambda^2}{2}-uv & \lambda u+u_x
 \\
 \lambda v-v_x & -\frac{\lambda^2}{2}+uv \\
 \end{array} \right).
 \label{lpt}
\end{eqnarray}
\label{lp}
\end{subequations}

Following the general procedure of NLP \cite{CaoGeng1990}, we consider $2N$ copies of the AKNS spectral problem (\ref{lpx}),
\begin{equation}
\left\{\begin{array}{l}
p_{j,x}=\frac{1}{2}\alpha_j p_j +u q_j,
\\
q_{j,x}=v p_j-\frac{1}{2}\alpha_j q_j,
\end{array}\right. 1\leq j\leq 2N,
\label{SPX}
\end{equation}
where $\alpha_j$, $1\leq j\leq 2N$, are $2N$ distinct eigenvalue parameters.
Consider the Bargmann constraint
\begin{equation}
u=-\langle \mathbf{p}, \mathbf{p} \rangle, \quad v=\langle \mathbf{q}, \mathbf{q}\rangle.
\label{con}
\end{equation}
Inserting the above constraint into (\ref{SPX}), we obtain the following finite-dimensional Hamiltonian
system
\begin{equation}
\left\{\begin{array}{l}
p_{j,x}= -\frac{\partial H_1}{\partial q_{j}}=\frac{1}{2}\alpha_j p_j - \langle \mathbf{p}, \mathbf{p} \rangle q_j,
\\
q_{j,x}=\frac{\partial H_1}{\partial p_{j}} =-\frac{1}{2}\alpha_j q_j +\langle \mathbf{q}, \mathbf{q} \rangle p_j,
\end{array}\right. 1\leq j\leq 2N,
\label{HX}
\end{equation}
where
\begin{eqnarray}
H_1=-\frac{1}{2}\langle \mathbf{\Lambda} \mathbf{p},\mathbf{q} \rangle+\frac{1}{2} \langle \mathbf{p},\mathbf{p} \rangle \langle \mathbf{q},\mathbf{q}\rangle,
\label{H1}
\end{eqnarray}
and $\mathbf{\Lambda}$ is defined by (\ref{notationpq}).
Moreover, the nonlinearized system corresponding to the time-part of the Lax system is
\begin{equation}\left\{\begin{array}{l}
p_{j,t}= -\frac{\partial H_2}{\partial q_{j}}=i\left(\frac{1}{2}\alpha_j^2+\langle \mathbf{p},\mathbf{p} \rangle \langle \mathbf{q},\mathbf{q}\rangle\right) p_j -i\left(\alpha_j \langle \mathbf{p}, \mathbf{p} \rangle+\langle \mathbf{\Lambda} \mathbf{p}, \mathbf{p} \rangle-2\langle \mathbf{p},\mathbf{p} \rangle \langle \mathbf{p},\mathbf{q}\rangle\right) q_j,
\\
q_{j,t}=\frac{\partial H_2}{\partial p_{j}}=i\left(\alpha_j \langle \mathbf{q}, \mathbf{q} \rangle+\langle \mathbf{\Lambda} \mathbf{q}, \mathbf{q} \rangle-2\langle \mathbf{q},\mathbf{q} \rangle \langle \mathbf{p},\mathbf{q}\rangle \right) p_j -i\left(\frac{1}{2}\alpha_j^2+\langle \mathbf{p},\mathbf{p} \rangle \langle \mathbf{q},\mathbf{q}\rangle\right) q_j,
\end{array}\right. 1\leq j\leq 2N,
\label{HT}
\end{equation}
 where
\begin{eqnarray}
H_2=-\frac{i}{2}\langle \mathbf{\Lambda}^2 \mathbf{p},\mathbf{q} \rangle +\frac{i}{2} \left(\langle \mathbf{\Lambda} \mathbf{p},\mathbf{p} \rangle\langle \mathbf{q},\mathbf{q} \rangle+\langle \mathbf{p},\mathbf{p} \rangle\langle \mathbf{\Lambda} \mathbf{q},\mathbf{q} \rangle\right)
-i\langle \mathbf{p},\mathbf{p} \rangle\langle \mathbf{q},\mathbf{q}\rangle\langle \mathbf{p},\mathbf{q} \rangle.
\label{H2}
\end{eqnarray}

We can check directly that the Hamiltonian systems (\ref{HX}) and (\ref{HT}) admit the Lax representations
\begin{eqnarray}
\frac{dL(\lambda)}{dx}=[\tilde{U},L(\lambda)],~~
\frac{dL(\lambda)}{dt}=[\tilde{V},L(\lambda)],
\end{eqnarray}
respectively, where $\tilde{U}$, $\tilde{V}$ are the matrices $U$, $V$ replacing $u$, $v$ and their derivatives by the Bargmann constraint, and the Lax matrix is given by
\begin{eqnarray}
L(\lambda)= \left( \begin{array}{cc} \frac{1}{2} & 0 \\
0 & -\frac{1}{2} \\ \end{array} \right)
+
\sum_{j=1}^{2N}\frac{1}{\lambda-\alpha_j}\left(
\begin{array}{cc} p_{j}q_{j} & -{p_{j}}^2\\ {q_{j}}^2 &
-p_{j}q_{j} \end{array}\right).
\label{LM}
\end{eqnarray}
Expanding $\det L(\lambda)$ as
\begin{equation}
F_{\lambda}= \det L(\lambda)=-\frac{1}{4}+\sum_{k=0}^\infty
F_k\lambda^{-k-1},
\label{F}
\end{equation}
we obtain
\begin{eqnarray}
\begin{split}
F_0=-\langle \mathbf{p},\mathbf{q}\rangle,
\\
F_1=-\langle \mathbf{\Lambda} \mathbf{p},\mathbf{q}\rangle+\langle \mathbf{p},\mathbf{p}\rangle\langle \mathbf{q},\mathbf{q}\rangle-\langle \mathbf{p},\mathbf{q}\rangle^2,
\\
F_k=-\langle \mathbf{\Lambda}^k \mathbf{p},\mathbf{q}\rangle+\sum_{l+j=k-1}\left(\langle \mathbf{\Lambda}^l \mathbf{p},\mathbf{p}\rangle\langle \mathbf{\Lambda}^j \mathbf{q},\mathbf{q}\rangle-\langle \mathbf{\Lambda}^l \mathbf{p},\mathbf{q}\rangle\langle \mathbf{\Lambda}^j \mathbf{p},\mathbf{q}\rangle\right), \quad k\geq 1.
\end{split}
\label{Fk}
\end{eqnarray}
We note that the polynomial integrals $\{F_k\}$ are usually called confocal system due to the connection of their generating function $F_{\lambda}$ with the theory of confocal quadrics \cite{Moser1980}.
By using the Poisson bracket (\ref{BP}), one can verify directly that the Lax matrix $L(\lambda)$ satisfies the following $r$-matrix relation,
\begin{eqnarray}
\left\{L_1(\lambda),L_2(\mu)\right\}=\left[r(\lambda,\mu),L_1(\lambda)+L_2(\mu)\right],
\label{rmatrix}
\end{eqnarray}
where $L_1(\lambda)=L(\lambda) \otimes \mathbf{I}$, $L_2(\mu)=\mathbf{I} \otimes L(\mu)$, $\mathbf{I}$ stands for the $2 \times 2$ identity matrix, and
\begin{eqnarray}
r(\lambda,\mu)=\frac{2}{\lambda-\mu}\left( \begin{array}{cccc}
1 & 0 & 0 & 0
\\
0 &  0 & 1 & 0
\\
0 & 1 & 0 & 0
\\
0 &  0 & 0 & 1
 \\ \end{array} \right).
 \label{rm}
\end{eqnarray}
According to the general theory of $r$-matrix theory \cite{BV1990,Faddeev2007}, we immediately obtain that the polynomial integrals $F_k$, $k\geq 0$, are involution in pair with respect to the Poisson bracket (\ref{BP}).
Moreover, it has been shown that $\left\{F_0, F_1,\cdots, F_{2N-1}\right\}$ is a set of functionally independent invariants in the open dense subset
$M=\{(\mathbf{p}^T,\mathbf{q}^T)\in \mathbb R^{4N}\mid(\mathbf{p}^T,\mathbf{q}^T)\neq0\}$ (see e.g. \cite{CaoGengWu1999}).
Thus, Hamiltonian systems (\ref{HX}) and (\ref{HT}) are completely integrable in the sense of Liouville since all the conditions of the Liouville-Arnold theorem are satisfied \cite{Moser1978,A1978}.

We note that the Hamiltonian $H_1$ and $H_2$ can be expressed in terms of $\{F_k\}$ as
\begin{eqnarray}
H_1=\frac{1}{2}\left(F_1+F_0^2\right)\equiv \tilde{H}_1, ~~ H_2=i\left(\frac{1}{2}F_2+F_1F_0+F_0^3\right)\equiv i\tilde{H}_2.
\end{eqnarray}
More generally, we may introduce the following set of polynomial integrals $\{\tilde{H}_k\}$,
\begin{eqnarray}
\begin{split}
\tilde{H}_0=\frac{1}{2}F_0,~~
\tilde{H}_{k+1}=\frac{1}{2}F_{k+1}+2\sum_{i+j=k}H_iH_j,~~k\geq 0,
\end{split}
\label{Hk}
\end{eqnarray}
which is equivalent to
\begin{equation}
(1-4H_{\lambda})^2=-4F_{\lambda},
\label{hamilton}
\end{equation}
where $H_{\lambda}=\sum_{k=0}^\infty \tilde{H}_k\lambda^{-k-1}$ is the generating function of $\{\tilde{H}_k\}$.
Sometimes, it is more convenient for us to deal with the generating function $H_{\lambda}$ than to deal with the Hamiltonian $H_1$ and $H_2$ separately.


\subsection{Linearization of the flows in the Jacobi variety}

Since the rational function $F_\lambda$ has simple poles at $\lambda=\alpha_1,\cdots,\alpha_{2N}$,
we may set
\begin{eqnarray}
F_{\lambda}= -\frac{1}{4}\frac{b(\lambda)}{\alpha(\lambda)}= -\frac{1}{4}\frac{R(\lambda)}{\alpha^2(\lambda)},
\label{FL}
\end{eqnarray}
where
\begin{eqnarray*}
\alpha(\lambda)= \prod_{j=1}^{2N}( \lambda-\alpha_j),~~b(\lambda)= \prod_{j=1}^{2N}( \lambda-\beta_j), ~~
R(\lambda)=\alpha(\lambda)b(\lambda)\equiv\prod_{j=1}^{4N}( \lambda-\alpha_j),
\end{eqnarray*}
with
$\beta_j\equiv \alpha_{2N+j},~~j=1,\cdots,2N$, being assumed to be distinct between each other.
Elliptic variables $ \mu_j$ and $ \nu_j$ are defined as the roots of $L^{12}({\lambda})$ and $L^{21}({\lambda})$, that is
\begin{eqnarray}
L^{12}({\lambda})=-\langle p,p\rangle \frac{m(\lambda)}{\alpha(\lambda)},
~~L^{21}({\lambda})=\langle q,q\rangle \frac{n(\lambda)}{\alpha(\lambda)},
\label{mu}
\end{eqnarray}
where
\begin{eqnarray*}
m(\lambda)=\prod_{j=1}^{2N-1}( \lambda-\mu_j),~~n(\lambda)=\prod_{j=1}^{2N-1}( \lambda-\nu_j).
\end{eqnarray*}
For fixed $\lambda_0$, we introduce the quasi-Abel-Jacobi variables:
\begin{equation}
\tilde \phi_s=\sum_{k=1}^g\int_{\lambda_0}^{\mu_k} \frac{\mu^{g-s}d\mu}{2\sqrt{R(\mu)}},
~~\tilde \psi_s=\sum_{k=1}^g\int_{\lambda_0}^{\nu_k} \frac{\nu^{g-s}d\nu}{2\sqrt{R(\nu)}},
~~ s=1,\cdots, g, ~~g=2N-1.\label{qaj}
\end{equation}
Following \cite{CaoGengWu1999}, we obtain
\begin{proposition}
The following relations hold
\begin{eqnarray}
\{\tilde \phi_s,F_{\lambda}\}=\frac{\lambda^{g-s}}{\alpha(\lambda)},
~~\{\tilde \psi_s,F_{\lambda}\}=-\frac{\lambda^{g-s}}{\alpha(\lambda)}, ~~s=1,\cdots,g, ~~g=2N-1.
\label{lamf}
\end{eqnarray}
\end{proposition}
By studying the expansion of $\frac{\lambda^{g}}{\alpha(\lambda)}$ in the negative powers of $\lambda$, we have
\begin{corollary}\label{qajs}
Let
\begin{equation}
S_l=\sum_{k=1}^{2N}\alpha_k^l,~~l\geq 1.
\label{S}
\end{equation}
Then
\begin{eqnarray}
\{\tilde \phi_s,F_{k}\}=A_{k-s}, ~~\{\tilde \psi_s,F_{k}\}=-A_{k-s},~~ s=1,\cdots,g, \quad
k=0,1\cdots.\label{tf}
\end{eqnarray}
where $A_{-j}=0$, $j\geq 1$, $A_0=1$, and $A_j$, $j\geq
1,$ are defined via $S_j$ by the following recursive formulae
\begin{equation}
A_1=S_1,~A_2=\frac{1}{2}\left(S_2+S_1^2\right),~A_k=\frac{1}{k}\left(S_k+\sum_{i+j=k,i,j\geq 1}S_iA_j\right),~k\geq 3.
\label{A}
\end{equation}
\end{corollary}

The formulae (\ref{qaj}) lead naturally to the consideration of the hyperelliptic curve $\Gamma$:
\begin{equation}
\zeta^2 =4R(\lambda)\equiv 4\prod_{j=1}^{4N}( \lambda-\alpha_j),
\label{cure1}
\end{equation}
with the genus $g=2N-1$.
The holomorphic differentials on $\Gamma$ are given by
$$\tilde
\omega_j=\frac{\lambda^{g-j}d\lambda}{\zeta},\quad
j=1,\cdots, g.$$
Since the degree of $R(\lambda)$ is even, there are two infinite points, denoted by $\infty_{+}$  and $\infty_{-}$,
which are located on the upper and lower sheets of $\Gamma$, respectively.
Let $a_1,\cdots,a_g, b_1,\cdots,b_g$ be canonical basis of the
homology group of cycles on the hyperelliptic curve
$\Gamma$. Denote by $A_{jk}$ the value of integral of
$\tilde \omega_j$ along $a_k$, and by $\mathbf{C}_1,\cdots,\mathbf{C}_g$ the column
vectors of the inverse matrix, i.e.
\begin{equation}
A_{jk}=\oint_{a_k} \tilde \omega_j, \quad \mathbf{C}=(A_{jk})^{-1}_{g\times
g}=(\mathbf{C}_1,\cdots,\mathbf{C}_g).
\end{equation}
For the normalized holomorphic differential $\omega_j$,
\begin{equation}
\boldsymbol{\omega}= \mathbf{C}\tilde{\boldsymbol{\omega}}, \quad \omega_s=\Sigma_{j=1}^g C_{sj} \tilde
\omega_j, \quad \tilde{\boldsymbol{\omega}}=(\tilde \omega_1,\cdots,\tilde
\omega_g)^T, \quad \boldsymbol{\omega}=( \omega_1,\cdots, \omega_g)^T,
\label{omega}
\end{equation}
we have
\begin{equation}
\oint_{a_k} \boldsymbol{\omega}=\boldsymbol{\delta}_{k}, \quad \oint_{b_k} \boldsymbol{\omega}=\mathbf{B}_{k}, \quad
k=1,\cdots,g,
\end{equation}
 where $(\boldsymbol{\delta}_{1},\cdots, \boldsymbol{\delta}_{g})$ is a unit matrix,
 and the matrix $(\mathbf{B}_{1},\cdots, \mathbf{B}_{g})\equiv \mathbf{B}$ is
 symmetric with positively definite imaginary part according to the Riemann bilinear relation \cite{GH1994}.
 The matrix $\mathbf{B}$ is used to construct the Riemann theta function \cite{M1982}:
\begin{equation}
\theta(\boldsymbol{\xi})\equiv\theta(\boldsymbol{\xi},\mathbf{B})=\sum_{\mathbf{z}\in\mathbb{Z}^g}\exp\left\{\pi i(\langle
\mathbf{B}\mathbf{z},\mathbf{z}\rangle+2\langle \boldsymbol{\xi},\mathbf{z}\rangle)\right\}, \quad \boldsymbol{\xi}\in\mathbb{C}^g,
\end{equation}
which has the following periodicity property
\begin{equation}
\begin{split}
\theta(\boldsymbol{\xi}+\mathbf{n})=\theta(\boldsymbol{\xi}), ~~\mathbf{n}\in \mathbb{Z}^g,
\\
\theta(\boldsymbol{\xi}+\mathbf{Bm})=\theta(\boldsymbol{\xi})\exp\left\{-\pi i\langle \mathbf{Bm},\mathbf{m}\rangle-2\pi i\langle \mathbf{m},\boldsymbol{\xi}\rangle\right\},~~\mathbf{m}\in \mathbb{Z}^g.
\end{split}
\label{thetapro}
\end{equation}
The periodic vectors $\boldsymbol{\delta}_{1},\cdots, \boldsymbol{\delta}_{g}, \mathbf{B}_{1},\cdots,
\mathbf{B}_{g}$ span a lattice $\mathcal{T}$ in $\mathbb{C}^g$, which defines
the Jacobi variety as a quotient manifold
$\mathcal{J}(\Gamma)=\mathbb{C}^g/ \mathcal{T}$. Let $Div(\Gamma)$
be the divisor group. The Abel
map $\mathcal{A}: Div(\Gamma)\rightarrow \mathcal{J}(\Gamma)$ is
defined as
\begin{equation}
\mathcal{A}(P)=\int_{P_0}^{P} \boldsymbol{\omega}, \qquad \mathcal{A}(\sum n_k
P_k)=\sum n_k \mathcal{A}(P_k),
\end{equation}
where $P_0$ is a fixed point on $\Gamma$.
The Abel-Jacobi variables $\phi_j$ and $\psi_j$ are defined as the normalization of
the quasi-Abel-Jacobi variables $\tilde \phi_j$ and $\tilde \psi_j$:
\begin{equation}
\mathbf{\Phi}=\mathbf{C} \tilde{\mathbf{\Phi}}= \mathcal{A}(\sum_{k=1}^{g} P(\mu_k)),
~~\mathbf{\Psi}=\mathbf{C} \tilde{\mathbf{\Psi}}= \mathcal{A}(\sum_{k=1}^{g} P(\nu_k)),
\label{AJC}
\end{equation}
where $\tilde{\mathbf{\Phi}}=(\tilde \phi_1,\cdots, \tilde\phi_g)^T$, $\mathbf{\Phi}=( \phi_1,\cdots, \phi_g)^T$, $\tilde{\mathbf{\Psi}}=(\tilde \psi_1,\cdots, \tilde\psi_g)^T$, $\mathbf{\Psi}=( \psi_1,\cdots, \psi_g)^T$.

Using (\ref{hamilton}), (\ref{lamf}) and (\ref{AJC}), we may deduce the following result.
\begin{proposition}\label{proAJ}
In the Jacobi variety,
\begin{equation}
\{\mathbf{\Phi},\tilde{H}_{k}\}=\boldsymbol{\Omega}_k, ~~\{\mathbf{\Psi},\tilde{H}_{k}\}=-\boldsymbol{\Omega}_k,
~~ k=0,1,\cdots,
\label{straigntout1}
\end{equation}
where
\begin{eqnarray}
\boldsymbol{\Omega}_0=0,~~\boldsymbol{\Omega}_1=\frac{1}{2}\mathbf{C}_1,
~~\boldsymbol{\Omega}_k=\frac{1}{2}(\Lambda_{k-1}\mathbf{C}_{1}+\cdots+\Lambda_{1}\mathbf{C}_{k-1}+\mathbf{C}_k),
~~k\geq 1,
\label{Omg}
\end{eqnarray}
with
\begin{eqnarray}
\Lambda_0=1,~~\Lambda_1=\frac{1}{2}\hat{S}_1,
~~\Lambda_k=\frac{1}{2k}\left(\hat{S}_{k}+\sum_{i+j=k,~i,j\geq 1}\hat{S}_{i}\Lambda_j\right),
~~k\geq 2,
\label{Lam}
\\
\hat{S}_k=\sum_{j=1}^{4N}\alpha_j^k=\sum_{j=1}^{2N}\alpha_j^k+\sum_{j=1}^{2N}\beta_j^k,~~k\geq 1.
\label{hatS}
\end{eqnarray}
\end{proposition}
Using (\ref{straigntout1}), we immediately obtain
\begin{proposition}\label{proAJHXT}
In the Jacobi variety, the Hamiltonian systems (\ref{HX}) and (\ref{HT}) are linearized respectively as
\begin{eqnarray}
\begin{split}
\frac{d \mathbf{\Phi}}{dx}=\boldsymbol{\Omega}_1, ~~\frac{d \mathbf{\Psi}}{dx}=-\boldsymbol{\Omega}_1,
\\
\frac{d \mathbf{\Phi}}{dt}=i\boldsymbol{\Omega}_2, ~~\frac{d \mathbf{\Psi}}{dt}=-i\boldsymbol{\Omega}_2,
\end{split}
\label{H12s}
\end{eqnarray}
where
\begin{eqnarray}
\boldsymbol{\Omega}_1=\frac{1}{2}\mathbf{C}_1,~~\boldsymbol{\Omega}_2=\frac{1}{2}(\Lambda_{1}\mathbf{C}_{1}+\mathbf{C}_2).
\label{om12}
\end{eqnarray}
\end{proposition}

\subsection{Application to the solutions of the AKNS equation}

The argument in section 2.1 implies the following result.
\begin{proposition} \label{cNLSdecom}
Let $\left(\mathbf{p}(x,t), \mathbf{q}(x,t)\right)$ be a compatible solution of the Hamiltonian systems (\ref{HX}) and (\ref{HT}).
Then (\ref{con}) solves the AKNS equation (\ref{akns}).
\end{proposition}

By virtue of (\ref{H12s}) and proposition \ref{cNLSdecom}, it is easy to obtain the solution of the AKNS equation in terms of the Abel-Jacobi coordinates
\begin{equation}
\mathbf{\Phi}(x,t)=\boldsymbol{\Omega}_1x + i\boldsymbol{\Omega}_2t+\mathbf{\Phi}_0,
~~\mathbf{\Psi}=-\boldsymbol{\Omega}_1x - i\boldsymbol{\Omega}_2t+\mathbf{\Phi}_0,
\label{sol}
\end{equation}
where $\mathbf{\Phi}_0\in \mathbb{C}^g$ is the vector defining the initial phase.
The solution of the AKNS equation in the original coordinates $u$ and $v$ can be derived via a standard inversion procedure \cite{GH1994,CaoGengWu1999}.
The resulting algebro-geometric solution is \cite{CaoGengWu1999}
\begin{eqnarray}
\begin{split}
u(x,t)=u(0,0)e^{N_1x+N_2t}\frac{\theta\left(\boldsymbol{\Omega}_1x+i\boldsymbol{\Omega}_2t+\mathbf{D}+\mathbf{\Delta}\right)
\theta\left(\mathbf{D}\right)}
{\theta\left(\boldsymbol{\Omega}_1x+i\boldsymbol{\Omega}_2t+\mathbf{D}\right)\theta\left(\mathbf{D}+\mathbf{\Delta}\right)},
\\
v(x,t)=v(0,0)e^{-N_1x-N_2t}\frac{\theta(\boldsymbol{\Omega}_1x+i\boldsymbol{\Omega}_2t-\mathbf{D}-\mathbf{\Delta})
\theta(\mathbf{D})}
{\theta(\boldsymbol{\Omega}_1x+i\boldsymbol{\Omega}_2t-\mathbf{D})\theta(\mathbf{D}+\mathbf{\Delta})},
\end{split}
\label{nlss1}
\end{eqnarray}
where $\boldsymbol{\Omega}_j$, $j=1,2$, are defined by (\ref{Omg}), and
\begin{subequations}
\begin{eqnarray}
N_1=\frac{1}{2}\hat{S}_1-I_1(\Gamma),~~I_s(\Gamma)=\sum_{j=1}^g \oint_{a_j}\lambda^s \omega_j,~~s=1,2,
\label{DDelta1}
\\
N_2=i\left(\frac{1}{2}\hat{S}_2-I_2(\Gamma)\right)
+2i\sum_{j,k=1}^g\Omega_{1}^j\Omega_{1}^k\partial_{jk}^2\ln\theta(\mathbf{D})
+2iu(0,0)v(0,0),
\label{DDelta2}
\\
\mathbf{D}=\mathbf{\Phi}_0+\mathbf{K}+\boldsymbol{\eta}_{+},
~~\boldsymbol{\eta}_{\pm}=\int_{\infty_{\pm}}^{P_0}\boldsymbol{\omega},
~~\mathbf{\Delta}=\int_{\gamma}\boldsymbol{\omega}=\int_{\infty_{-}}^{\infty_{+}}\boldsymbol{\omega},
\label{DDelta3}
\end{eqnarray}
\label{DDelta}
\end{subequations}
and $\mathbf{K}$ is the vector of Riemann constants, $\gamma$ is the path connecting the points $\infty_{\pm}$ and intersecting none of the $a$-cycles, $\Omega_{1}^j$ denotes the $j$-th element of the vector $\boldsymbol{\Omega}_1$,
and the notation $\partial_{jk}^2 f(\boldsymbol{\xi})$ stands for $\frac{\partial^2}{\partial\xi_j\partial\xi_k}f(\boldsymbol{\xi})$.
We note that the algebro-geometric solution of the AKNS equation in the form of (\ref{nlss1}) coincides with the one presented in \cite{BBEIM1994}.

\section{NLP of the AKNS equation with space-inverse reductions}

In this section, we show how to deal with the NLP method that permits to impose space-inverse reductions on the AKNS equation. As a consequence, we obtain a new type of finite-dimensional integrable Hamiltonian systems which is nonlocal with respect to the space variable. These new nonlocal Hamiltonian systems constitute finite-dimensional decompositions of the AKNS equation in the presence of space-inverse reductions.

\subsection{NLP of the AKNS equation with odd/even reduction (\ref{oddevenuv})}

\begin{lemma}\label{lemma1}
Suppose that the reduction (\ref{oddevenuv}) holds. If $(p(x,t),q(x,t))^T$ solves (\ref{lp}) with parameter $\lambda$, then $\sqrt{\epsilon} \left(p(-x,t),-\epsilon q(-x,t)\right)^T$ solves (\ref{lp}) with parameter $-\lambda$.
\end{lemma}

The above observation suggests us to consider the following paired eigenvalue parameters and eigenfunctions in the standard procedure of NLP,
\begin{subequations}
\begin{eqnarray}
\alpha_j,~\alpha_{N+j}=-\alpha_{j},~1\leq j\leq N,
\label{DRDa}
\\
\left(p_{j}(x,t),q_{j}(x,t)\right)^T,~\left(p_{N+j}(x,t),q_{N+j}(x,t)\right)^T=\sqrt{\epsilon} \left(p_{j}(-x,t),-\epsilon q_{j}(-x,t)\right)^T,~1\leq j\leq N.
\label{DRDb}
\end{eqnarray}
\label{DRD}
\end{subequations}
For convenience sake, we will use the following notations in what follows
\begin{eqnarray}
\begin{split}
\mathbf{A}=diag(\alpha_1,\alpha_2,\cdots,\alpha_{N}),
\\
\mathbf{P}\equiv(p_{1},p_{2},\cdots p_{N})^T=(p_{1}(x,t),p_{2}(x,t),\cdots p_{N}(x,t))^T,
\\
\hat{\mathbf{P}}\equiv(\hat{p}_{1},\hat{p}_{2},\cdots \hat{p}_{N})^T=(p_{1}(-x,t),p_{2}(-x,t),\cdots p_{N}(-x,t))^T,
\\
\mathbf{Q}\equiv (q_{1},q_{2},\cdots q_{N})^T=(q_{1}(x,t),q_{2}(x,t),\cdots q_{N}(x,t))^T,
\\
\hat{\mathbf{Q} }\equiv(\hat{q}_{1},\hat{q}_{2},\cdots \hat{q}_{N})^T=(q_{1}(-x,t),q_{2}(-x,t),\cdots q_{N}(-x,t))^T.
\end{split}
\label{notationpq1}
\end{eqnarray}
Using (\ref{DRD}), the constraint (\ref{con}) becomes
\begin{eqnarray}
\begin{split}
u(x,t)=-\langle \mathbf{P},\mathbf{P}\rangle-\epsilon\langle \hat{\mathbf{P}},\hat{\mathbf{P}}\rangle,
~~
v(x,t)=\langle \mathbf{Q},\mathbf{Q}\rangle+\epsilon\langle \hat{\mathbf{Q}},\hat{\mathbf{Q}}\rangle,
\end{split}
\label{uvpq1}
\end{eqnarray}
and the finite-dimensional Hamiltonian system associated with the space-part of the Lax pair becomes
\begin{equation}\left\{\begin{array}{l}
p_{j,x}=\frac{1}{2}\alpha_j p_j -\left( \langle \mathbf{P}, \mathbf{P}\rangle+\epsilon\langle \hat{\mathbf{P}}, \hat{\mathbf{P}} \rangle\right)q_j= -\frac{\partial H_1}{\partial q_{j}},
\\
q_{j,x}=-\frac{1}{2}\alpha_j q_j +\left( \langle \mathbf{Q},\mathbf{Q}\rangle+\epsilon\langle \hat{\mathbf{Q}},\hat{\mathbf{Q}}\rangle\right) p_j =\frac{\partial H_1}{\partial p_{j}},
\\
\hat{p}_{j,x}=-\frac{1}{2}\alpha_j \hat{p}_j +\left( \epsilon\langle \mathbf{P}, \mathbf{P}\rangle+\langle \hat{\mathbf{P}}, \hat{\mathbf{P}} \rangle\right)\hat{q}_j= \frac{\partial H_1}{\partial \hat{q}_{j}},
\\
\hat{q}_{j,x}=\frac{1}{2}\alpha_j \hat{q}_j -\left( \epsilon\langle \mathbf{Q},\mathbf{Q}\rangle+\langle \hat{\mathbf{Q}},\hat{\mathbf{Q}}\rangle\right) \hat{p}_j =-\frac{\partial H_1}{\partial \hat{p}_{j}} ,
\end{array}\right. 1\leq j\leq N,
\label{HXD}
\end{equation}
where
\begin{eqnarray}
\begin{split}
H_1=&-\frac{1}{2}\left(\langle \mathbf{A} \mathbf{P},\mathbf{Q} \rangle+\langle \mathbf{A} \hat{\mathbf{P}},\hat{\mathbf{Q}} \rangle\right)
+\frac{1}{2} \left(\langle \mathbf{P},\mathbf{P}\rangle+\epsilon\langle \hat{\mathbf{P}},\hat{\mathbf{P}} \rangle\right)
\left(\langle \mathbf{Q},\mathbf{Q}\rangle+\epsilon\langle \hat{\mathbf{Q}},\hat{\mathbf{Q}} \rangle\right).
\end{split}
\label{H1D}
\end{eqnarray}
The finite-dimensional Hamiltonian system associated with the time-part of Lax pair becomes
\begin{equation}\left\{\begin{array}{l}
p_{j,t}=-\frac{\partial H_2}{\partial q_{j}},
\\
q_{j,t}=\frac{\partial H_2}{\partial p_{j}},
\\
\hat{p}_{j,t}=\frac{\partial H_2}{\partial \hat{q}_{j}},
\\
\hat{q}_{j,t}=-\frac{\partial H_2}{\partial \hat{p}_{j}} ,
\end{array}\right. 1\leq j\leq N,
\label{HTD}
\end{equation}
 where
\begin{eqnarray}
\begin{split}
H_2=&-\frac{i}{2}\left(\langle \mathbf{A}^2 \mathbf{P},\mathbf{Q} \rangle-\langle \mathbf{A}^2 \hat{\mathbf{P}},\hat{\mathbf{Q}} \rangle\right)
\\&+\frac{i}{2} \left(\langle \mathbf{A} \mathbf{P},\mathbf{P} \rangle-\epsilon\langle \mathbf{A} \hat{\mathbf{P}},\hat{\mathbf{P}} \rangle\right)\left(\langle \mathbf{Q},\mathbf{Q} \rangle+\epsilon\langle \hat{\mathbf{Q}},\hat{\mathbf{Q}} \rangle\right)
\\&+\frac{i}{2} \left(\langle\mathbf{P},\mathbf{P} \rangle+\epsilon\langle\hat{\mathbf{P}},\hat{\mathbf{P}} \rangle\right)\left(\langle \mathbf{A}\mathbf{Q},\mathbf{Q} \rangle-\epsilon\langle \mathbf{A}\hat{\mathbf{Q}},\hat{\mathbf{Q}} \rangle\right)
\\&-i\left(\langle\mathbf{P},\mathbf{P} \rangle+\epsilon\langle\hat{\mathbf{P}},\hat{\mathbf{P}} \rangle\right)\left(\langle\mathbf{Q},\mathbf{Q} \rangle+\epsilon\langle\hat{\mathbf{Q}},\hat{\mathbf{Q}} \rangle\right)
\left(\langle \mathbf{P},\mathbf{Q} \rangle-\langle \hat{\mathbf{P}},\hat{\mathbf{Q}}\rangle\right).
\end{split}
\label{H2D}
\end{eqnarray}
We can verify directly that the last two equations of (\ref{HXD}) are consistent with the first two equations of (\ref{HXD}), and so does (\ref{HTD}). Thus the Hamiltonian systems (\ref{HX}) and (\ref{HT}) do admit the nonlocal reduction (\ref{DRD}).

Motivated by (\ref{symstruce}) and (\ref{DRD}), we introduce the symplectic structure
\begin{equation}
\omega=\sum_{j=1}^N \left(dp_j\wedge dq_j-d\hat{p}_j\wedge d\hat{q}_j\right).
\label{symp1}
\end{equation}
The corresponding Poisson bracket reads
\begin{equation}
\left\{f,g\right\}=\sum_{j=1}^N\left(\frac{\partial f}{\partial
q_j} \frac{\partial g}{\partial p_j}-\frac{\partial f}{\partial p_j}
\frac{\partial g}{\partial q_j}-\frac{\partial f}{\partial
\hat{q}_j} \frac{\partial g}{\partial \hat{p}_j}+\frac{\partial f}{\partial \hat{p}_j}
\frac{\partial g}{\partial \hat{q}_j}\right).
\label{BP1}
\end{equation}
Using (\ref{DRD}), the Lax matrix (\ref{LM}) becomes
\begin{eqnarray}
\begin{split}
L(\lambda)=& \left( \begin{array}{cc} \frac{1}{2} & 0 \\
0 & -\frac{1}{2} \\ \end{array} \right)
+
\sum_{j=1}^{N}\frac{1}{\lambda-\alpha_j}\left(
\begin{array}{cc} p_{j}q_{j} & -{p_{j}}^2\\ {q_{j}}^2 &
-p_{j}q_{j}\end{array}\right)
+\sum_{j=1}^{N}\frac{1}{\lambda+\alpha_j}\left(
\begin{array}{cc} -\hat{p}_{j}\hat{q}_{j} & -\epsilon{\hat{p}_{j}}^2\\ \epsilon{\hat{q}_{j}}^2 &
\hat{p}_{j}\hat{q}_{j} \end{array}\right),
\end{split}
\label{LMD}
\end{eqnarray}
whose determinant can be expanded as
\begin{equation}
\det L(\lambda)=-\frac{1}{4}+\sum_{k=0}^\infty
F_k\lambda^{-k-1},
\label{FD}
\end{equation}
where
\begin{eqnarray*}
\begin{split}
F_0=&-\langle \mathbf{P},\mathbf{Q}\rangle+\langle \hat{\mathbf{P}},\hat{\mathbf{Q}}\rangle,
\\
F_1=&-F^2_0-\langle \mathbf{A} \mathbf{P},\mathbf{Q}\rangle-\langle \mathbf{A} \hat{\mathbf{P}},\hat{\mathbf{Q}} \rangle
+\left(\langle \mathbf{P},\mathbf{P} \rangle+\epsilon\langle \hat{\mathbf{P}},\hat{\mathbf{P}}\rangle\right)
\left(\langle \mathbf{Q},\mathbf{Q}\rangle+\epsilon\langle \hat{\mathbf{Q}},\hat{\mathbf{Q}}\rangle\right),
\\
F_k=&-\langle \mathbf{A}^k \mathbf{P},\mathbf{Q}\rangle+(-1)^k\langle \mathbf{A}^k \hat{\mathbf{P}},\hat{\mathbf{Q}}\rangle
\\&+\sum_{l+j=k-1}\{(\langle \mathbf{A}^l \mathbf{P},\mathbf{P}\rangle+\epsilon\langle (-\mathbf{A})^l \hat{\mathbf{P}},\hat{\mathbf{P}}\rangle)(\langle \mathbf{A}^j \mathbf{Q},\mathbf{Q}\rangle+\epsilon\langle (-\mathbf{A})^j \hat{\mathbf{Q}},\hat{\mathbf{Q}}\rangle)
\\&
-(\langle \mathbf{A}^l \mathbf{P},\mathbf{Q}\rangle-\langle (-\mathbf{A})^l \hat{\mathbf{P}},\hat{\mathbf{Q}}\rangle)(\langle \mathbf{A}^j \mathbf{P},\mathbf{Q}\rangle-\langle (-\mathbf{A})^j \hat{\mathbf{P}},\hat{\mathbf{Q}}\rangle)\},~~k\geq 2.
\end{split}
\label{FkD}
\end{eqnarray*}
Based on the Poisson bracket (\ref{BP1}), we may verify directly that the Lax matrix (\ref{LMD}) satisfies the same $r$-matrix relation as the one for the Lax matrix (\ref{LM}) (see (\ref{rmatrix})).
This fact implies that the polynomial integrals $\{F_k\}_{k\geq 0}$ are involutive in pairs with respect to the Poisson bracket (\ref{BP1}).
Moreover, we can show, as in \cite{CaoGengWu1999}, that the integrals $\left\{F_0, F_1,\cdots, F_{2N-1}\right\}$ are functionally independent in the open dense subset
$M=\{(\mathbf{P}^T,\mathbf{Q}^T,\hat{\mathbf{P}}^T,\hat{\mathbf{Q}}^T)\in \mathbb R^{4N}\mid(\mathbf{P}^T,\mathbf{Q}^T,\hat{\mathbf{P}}^T,\hat{\mathbf{Q}}^T)\neq0\}$.
Thus, the nonlocal Hamilton systems (\ref{HXD}) and (\ref{HTD}) are completely integrable in the sense of Liouville.
Here we note that in the theory of classical integrable Hamiltonian systems, the symplectic structure on the phase space touch no information about the independent flow variable $x$. Thus in order to make our results more easily to be understood within the framework of classical integrable Hamiltonian systems, we here prefer to view $\hat{\mathbf{P}}$, $\hat{\mathbf{Q}}$ as independent variables from $\mathbf{P}$, $\mathbf{Q}$ in the phase space, despite the fact that they may relate to each other when they are considered as the functions of the flow variable $x$.

The above analysis implies that
\begin{proposition}\label{prouvs1}
Let $\left(\mathbf{P}(x,t), \mathbf{Q}(x,t)\right)$ be a compatible solution of the nonlocal Hamiltonian systems (\ref{HXD}) and (\ref{HTD}).
Then $\left(u,v\right)$, constructed by (\ref{uvpq1}), solves the AKNS equation (\ref{akns}) with the space-inverse symmetry (\ref{oddevenuv}). In other words, (\ref{HXD}) and (\ref{HTD}) constitute a finite-dimensional decomposition of the AKNS equation (\ref{akns}) in the presence of the space-inverse symmetry (\ref{oddevenuv}).
\end{proposition}

We now study the linearization of the nonlocal Hamiltonian systems (\ref{HXD}) and (\ref{HTD}) in the Jacobi variety.
For the Lax matrix (\ref{LMD}), the rational functions $F(\lambda)=\det L(\lambda)$, $L^{12}({\lambda})$ and $L^{21}({\lambda})$ have simple poles at $\lambda=\pm\alpha_j$, $1\leq j\leq N$.
It is easy to verify that
\begin{eqnarray}
L(\lambda,x)=\left( \begin{array}{cc} -\epsilon & 0 \\
0 & 1 \\ \end{array} \right)L(-\lambda,-x)\left( \begin{array}{cc} -\epsilon & 0 \\
0 & 1 \\ \end{array} \right).
\label{lxsym}
\end{eqnarray}
Using (\ref{lxsym}) and the fact that $F(\lambda)$ is invariant along the $H_1$-flow, we obtain the symmetry $F(\lambda)=F(-\lambda)$.
This symmetry property implies that along the $H_1$-flow the zeros of $F(\lambda)$ always appear in opposite pairs. We denote such pairs as $\left\{\pm\beta_j\right\}_1^N$.
Thus we may set
\begin{subequations}
\begin{eqnarray}
F(\lambda)= -\frac{1}{4}\frac{b(\lambda)}{\alpha(\lambda)}= -\frac{1}{4}\frac{R(\lambda)}{\alpha^2(\lambda)},\label{FLDa}
\\
L^{12}({\lambda})=-\left(\langle \mathbf{P},\mathbf{P}\rangle+\epsilon\langle \hat{\mathbf{P}},\hat{\mathbf{P}}\rangle\right) \frac{m(\lambda)}{\alpha(\lambda)},\label{FLDb}
\\
L^{21}({\lambda})=\left(\langle \mathbf{Q},\mathbf{Q}\rangle+\epsilon\langle \hat{\mathbf{Q}},\hat{\mathbf{Q}}\rangle \right)\frac{n(\lambda)}{\alpha(\lambda)},\label{FLDc}
\end{eqnarray}
\label{FLD}
\end{subequations}
where
\begin{subequations}
\begin{eqnarray}
\alpha(\lambda)= \prod_{j=1}^{N}( \lambda^2-\alpha_j^2),~~b(\lambda)= \prod_{j=1}^{N}( \lambda^2-\beta^2_j),\label{RmnDa}
\\
m(\lambda)=\prod_{j=1}^{2N-1}( \lambda-\mu_j),~~n(\lambda)=\prod_{j=1}^{2N-1}( \lambda-\nu_j),\label{RmnDb}
\\
R(\lambda)=\alpha(\lambda)b(\lambda)\equiv\prod_{j=1}^{2N}( \lambda^2-\lambda^2_j),~~\lambda_j=\alpha_j,~~\lambda_{N+j}=\beta_j,~~j=1,\cdots,N. \label{RmnDc}
\end{eqnarray}
\label{RmnD}
\end{subequations}
We define the quasi-Abel-Jacobi variables $\tilde\phi_s$, $\tilde\psi_s$, $1\leq s\leq 2N-1$, in the same form as (\ref{qaj}) but with the functions $R(\lambda)$, $\mu_j$ and $\nu_j$ defined by (\ref{FLD}) and (\ref{RmnD}) instead of (\ref{FL}) and (\ref{mu}).
The formula (\ref{RmnDc}) implies the consideration of the following hyperelliptic curve $\Gamma$ of genus $g=2N-1$,
\begin{eqnarray}
\zeta^2 =4\prod_{j=1}^{2N}( \lambda^2-\lambda^2_j)\equiv 4\prod_{j=1}^{4N}(\lambda-\lambda_j),
~~\lambda_{2N+k}=-\lambda_k,~~1\leq k\leq 2N,
\label{HCD}
\end{eqnarray}
where $\lambda_j\neq \lambda_l$, $j\neq l$, $1\leq j, l\leq 4N$.
Let $a_1,\cdots,a_g$, $b_1,\cdots,b_g$ be canonical basis of cycles on $\Gamma$.
Denote by $A_{jk}$ the value of integral of the holomorphic differential $\tilde \omega_j$ along $a_k$, and denote by $\mathbf{C}_j$ the column vectors of the inverse matrix $\mathbf{C}=\left(\int_{a_k}\tilde{\omega}_j\right)^{-1}_{g\times g}$.
As in section 2.2, we define the Abel-Jacobi variables $\mathbf{\Phi}=(\phi_1,\cdots,\phi_g)^T$ and $\mathbf{\Psi}=(\psi_1,\cdots, \psi_g)^T$ as the normalization of
the quasi-Abel-Jacobi variables $\tilde{\mathbf{\Phi}}=(\tilde \phi_1,\cdots, \tilde\phi_g)^T$ and $\tilde{ \mathbf{\Psi}}=(\tilde \psi_1,\cdots, \tilde\psi_g)^T$, that is
\begin{equation}
\mathbf{\Phi}=\left(\mathbf{C}_1,\cdots,\mathbf{C}_g\right)\tilde{\mathbf{\Phi}}=\mathcal{A}(\sum_{k=1}^{g} P(\mu_k)),
~~\mathbf{\Psi}=\left(\mathbf{C}_1,\cdots,\mathbf{C}_g\right)\tilde{\mathbf{\Psi}}=\mathcal{A}(\sum_{k=1}^{g} P(\nu_k)).\label{AJD}
\end{equation}

Inserting the symmetries $\alpha_{N+j}=-\alpha_j$, $\beta_{N+j}=-\beta_j$, $j=1,\cdots,N$, into (\ref{Omg}), (\ref{Lam}) and (\ref{hatS}), we obtain
\begin{subequations}
\begin{eqnarray}
\hat{S}_{2k-1}=0,~~\hat{S}_{2k}=2\sum_{j=1}^{2N}\lambda_j^{2k},~~k\geq 1,
\label{hatsD}
\\
\Lambda_0=1,~~\Lambda_{2k-1}=0,
~~\Lambda_{2k}=\frac{1}{4k}\left(\hat{S}_{2k}+\sum_{i+j=2k,~i,j\geq 1}\hat{S}_{i}\Lambda_j\right),
~~k\geq 1,
\label{LamD}
\\
\boldsymbol{\Omega}_0=0,~~\boldsymbol{\Omega}_1=\frac{1}{2}\mathbf{C}_1,
~~\boldsymbol{\Omega}_2=\frac{1}{2}\mathbf{C}_2,
~~\boldsymbol{\Omega}_k=\frac{1}{2}\sum_{i+j=k,~i\geq 0,~j\geq 1}\Lambda_{i}\mathbf{C}_{j},
~~k\geq 3
\label{OmgD}
\end{eqnarray}
\label{sloD}
\end{subequations}
By using the formulae (\ref{sloD}) and the proposition \ref{proAJHXT}, we immediately obtain
\begin{proposition}
In the Jacobi variety, the nonlocal Hamiltonian systems (\ref{HXD}) and (\ref{HTD}) are linearized as
\begin{eqnarray}
\begin{split}
\frac{d \mathbf{\Phi}}{dx}=\frac{1}{2}\mathbf{C}_1, ~~\frac{d \mathbf{\Psi}}{dx}=-\frac{1}{2}\mathbf{C}_1,
\\
\frac{d \mathbf{\Phi}}{dt}=\frac{i}{2}\mathbf{C}_2, ~~\frac{d \mathbf{\Psi}}{dt}=-\frac{i}{2}\mathbf{C}_2.
\end{split}
\label{H12sD}
\end{eqnarray}
\end{proposition}

\begin{remark}
The hyperelliptic curve (\ref{HCD}), compared to (\ref{cure1}), admits an additional holomorphic involution. This involution enables us to construct algebro-geometric solutions to the AKNS equation in the presence of the Dirichlet/Neumann boundary conditions. This issue will be discussed in details in section 4.1.
\end{remark}

\subsection{NLP of the AKNS equation with nonlocal reduction (\ref{nonlocalred})}

\begin{lemma}\label{lemma1}
Suppose that the reduction (\ref{nonlocalred}) holds. If $(p(x,t),q(x,t))^T$ solves (\ref{lp}) with parameter $\lambda$, then $\sqrt{-\epsilon} \left(\bar{q}(-x,t),-\epsilon \bar{p}(-x,t)\right)^T$ solves (\ref{lp}) with parameter $\bar{\lambda}$.
\end{lemma}

The above observation suggests us to consider the following paired eigenvalue parameters and eigenfunctions in the standard procedure of NLP,
\begin{eqnarray}
\begin{split}
\alpha_j,~\alpha_{N+j}=\bar{\alpha}_{j},~1\leq j\leq N,
\\
\left(p_{j}(x,t),q_{j}(x,t)\right)^T,~\left(p_{N+j}(x,t),q_{N+j}(x,t)\right)^T=\sqrt{-\epsilon} \left(\bar{q}_{j}(-x,t),-\epsilon \bar{p}_{j}(-x,t)\right)^T,~1\leq j\leq N.
\end{split}
\label{DRnol}
\end{eqnarray}
Using (\ref{DRnol}), the constraint (\ref{con}) becomes
\begin{eqnarray}
\begin{split}
u(x,t)=-\langle \mathbf{P},\mathbf{P}\rangle+\epsilon\overline{\langle \hat{\mathbf{Q}},\hat{\mathbf{Q}}\rangle},
~~
v(x,t)=\langle \mathbf{Q},\mathbf{Q}\rangle-\epsilon\overline{\langle \hat{\mathbf{P}},\hat{\mathbf{P}}\rangle},
\end{split}
\label{uvpqnol}
\end{eqnarray}
where $\mathbf{P}$, $\mathbf{Q}$, $\hat{\mathbf{P}}$ and $\hat{\mathbf{Q}}$ are the notations introduced in (\ref{notationpq}).
The finite-dimensional Hamiltonian system associated with the space-part of the Lax pair becomes
\begin{equation}\left\{\begin{array}{l}
p_{j,x}=\frac{1}{2}\alpha_j p_j -\left( \langle \mathbf{P}, \mathbf{P}\rangle-\epsilon\overline{\langle \hat{\mathbf{Q}},\hat{\mathbf{Q}}\rangle}\right)q_j= -\frac{\partial H_1}{\partial q_{j}},
\\
q_{j,x}=-\frac{1}{2}\alpha_j q_j +\left( \langle \mathbf{Q},\mathbf{Q}\rangle-\epsilon\overline{\langle \hat{\mathbf{P}},\hat{\mathbf{P}}\rangle}\right) p_j =\frac{\partial H_1}{\partial p_{j}},
\\
\bar{\hat{q}}_{j,x}=\frac{1}{2}\bar{\alpha}_j \bar{\hat{q}}_j +\left( \epsilon\langle \mathbf{P}, \mathbf{P}\rangle-\overline{\langle \hat{\mathbf{Q}}, \hat{\mathbf{Q}} \rangle}\right)\bar{\hat{p}}_j
=-\frac{\partial H_1}{\partial \bar{\hat{p}}_{j}},
\\
\bar{\hat{p}}_{j,x}=-\frac{1}{2}\bar{\alpha}_j \bar{\hat{p}}_j
+\left(\overline{\langle \hat{\mathbf{P}},\hat{\mathbf{P}}\rangle}
-\epsilon\langle \mathbf{Q},\mathbf{Q}\rangle\right) \bar{\hat{q}}_j
=\frac{\partial H_1}{\partial \bar{\hat{q}}_{j}} ,
\end{array}\right. 1\leq j\leq N,
\label{HXnol}
\end{equation}
where
\begin{eqnarray}
H_1=-\frac{1}{2}\left(\langle \mathbf{A} \mathbf{P},\mathbf{Q}\rangle+\overline{\langle \mathbf{A} \hat{\mathbf{P}},\hat{\mathbf{Q}}\rangle}\right)
+\frac{1}{2} \left(\langle \mathbf{P},\mathbf{P}\rangle-\epsilon\overline{\langle \hat{\mathbf{Q}},\hat{\mathbf{Q}}\rangle}\right)
\left(\langle \mathbf{Q},\mathbf{Q}\rangle-\epsilon\overline{\langle \hat{\mathbf{P}},\hat{\mathbf{P}}\rangle}\right).
\label{H1nol}
\end{eqnarray}
The finite-dimensional Hamiltonian system associated with the time-part of Lax pair becomes
\begin{equation}\left\{\begin{array}{l}
p_{j,t}=-\frac{\partial H_2}{\partial q_{j}},
\\
q_{j,t}=\frac{\partial H_2}{\partial p_{j}},
\\
\bar{\hat{q}}_{j,t}=-\frac{\partial H_2}{\partial \bar{\hat{p}}_{j}},
\\
\bar{\hat{p}}_{j,t}=\frac{\partial H_2}{\partial \bar{\hat{q}}_{j}} ,
\end{array}\right. 1\leq j\leq N,
\label{HTnol}
\end{equation}
 where
\begin{eqnarray}
\begin{split}
H_2=&-\frac{i}{2}\left(\langle \mathbf{A}^2 \mathbf{P},\mathbf{Q} \rangle+\overline{\langle \mathbf{A}^2 \hat{\mathbf{P}},\hat{\mathbf{Q}} \rangle}\right)
\\&+\frac{i}{2} \left(\langle \mathbf{A} \mathbf{P},\mathbf{P} \rangle-\epsilon\overline{\langle \mathbf{A} \hat{\mathbf{Q}},\hat{\mathbf{Q}} \rangle}\right)\left(\langle \mathbf{Q},\mathbf{Q} \rangle-\epsilon\overline{\langle \hat{\mathbf{P}},\hat{\mathbf{P}} \rangle}\right)
\\&+\frac{i}{2} \left(\langle\mathbf{P},\mathbf{P} \rangle-\epsilon\overline{\langle\hat{\mathbf{Q}},\hat{\mathbf{Q}} \rangle}\right)\left(\langle \mathbf{A}\mathbf{Q},\mathbf{Q} \rangle-\epsilon\overline{\langle \mathbf{A}\hat{\mathbf{P}},\hat{\mathbf{P}} \rangle}\right)
\\&-i\left(\langle\mathbf{P},\mathbf{P} \rangle-\epsilon\overline{\langle\hat{\mathbf{Q}},\hat{\mathbf{Q}} \rangle}\right)\left(\langle\mathbf{Q},\mathbf{Q} \rangle-\epsilon\overline{\langle\hat{\mathbf{P}},\hat{\mathbf{P}} \rangle}\right)
\left(\langle \mathbf{P},\mathbf{Q} \rangle+\overline{\langle \hat{\mathbf{P}},\hat{\mathbf{Q}}\rangle}\right).
\end{split}
\label{H2nol}
\end{eqnarray}
Again we can verify directly that the last two equations of (\ref{HXnol}) are consistent with the first two equations of (\ref{HXnol}), and so does (\ref{HTnol}). Thus the Hamiltonian systems (\ref{HXnol}) and (\ref{HTnol}) do admit the nonlocal reduction (\ref{DRnol}).

Inspired by (\ref{DRnol}), we introduce the symplectic structure
\begin{equation}
\omega=\sum_{j=1}^N \left(dp_j\wedge dq_j+d\bar{\hat{q}}_j\wedge d\bar{\hat{p}}_j\right).
\label{symp2}
\end{equation}
The corresponding Poisson bracket is given by
\begin{equation}
\left\{f,g\right\}=\sum_{j=1}^N\left(\frac{\partial f}{\partial
q_j} \frac{\partial g}{\partial p_j}-\frac{\partial f}{\partial p_j}
\frac{\partial g}{\partial q_j}+\frac{\partial f}{\partial \bar{\hat{p}}_j}
\frac{\partial g}{\partial \bar{\hat{q}}_j}-\frac{\partial f}{\partial
\bar{\hat{q}}_j} \frac{\partial g}{\partial \bar{\hat{p}}_j}\right).
\label{BP2}
\end{equation}
The Lax matrix associated with the reduction (\ref{DRnol}) becomes
\begin{eqnarray}
\begin{split}
L(\lambda)=& \left( \begin{array}{cc} \frac{1}{2} & 0 \\
0 & -\frac{1}{2} \\ \end{array} \right)
+
\sum_{j=1}^{N}\frac{1}{\lambda-\alpha_j}\left(
\begin{array}{cc} p_{j}q_{j} & -{p_{j}}^2\\ {q_{j}}^2 &
-p_{j}q_{j}\end{array}\right)
+\sum_{j=1}^{N}\frac{1}{\lambda-\bar{\alpha}_j}\left(
\begin{array}{cc} \bar{\hat{p}}_{j}\bar{\hat{q}}_{j} & \epsilon{\bar{\hat{q}}_{j}}^2\\
-\epsilon{\bar{\hat{p}}_{j}}^2 &-\bar{\hat{p}}_{j}\bar{\hat{q}}_{j} \end{array}\right).
\end{split}
\label{LM3}
\end{eqnarray}
Using the Poisson bracket (\ref{BP2}), we may verify directly that the Lax matrix (\ref{LM3}) satisfies the $r$-matrix relation (\ref{rmatrix}).
As in section 3.1, we can extract functionally independent and involutive pairwise integrals of motion $\left\{F_0, F_1, \cdots, F_{2(N-1)}\right\}$ by considering the expansion of $F({\lambda})\equiv\det L(\lambda)$ in powers of $\lambda^{-1}$.
Thus, our nonlocal Hamiltonian systems (\ref{HXnol}) and (\ref{HTnol}) are integrable in the sense of Liouville.

The above arguments imply that
\begin{proposition}\label{prouvs3}
Let $\left(\mathbf{P}(x,t), \mathbf{Q}(x,t)\right)$ be a compatible solution of the nonlocal Hamiltonian systems (\ref{HXnol}) and (\ref{HTnol}). Then $u(x,t)$, constructed by (\ref{uvpqnol}), solves the nonlocal NLS equation (\ref{nnls}).
In other words, (\ref{HXnol}) and (\ref{HTnol}) constitute a finite-dimensional decomposition of the nonlocal NLS equation (\ref{nnls}).
\end{proposition}

We now consider the linearization of the nonlocal Hamiltonian systems (\ref{HXnol}) and (\ref{HTnol}) in the Jacobi variety.
It is easy to check that (\ref{LM3}) satisfies the symmetry
\begin{eqnarray}
L(\lambda,x)=\left( \begin{array}{cc} 0 & \epsilon \\
-1 & 0 \\ \end{array} \right)
\overline{L(\bar{\lambda},-x)}
\left( \begin{array}{cc} 0 & 1 \\
-\epsilon & 0 \\ \end{array} \right).
\label{symL}
\end{eqnarray}
Using (\ref{symL}) and the fact that $F({\lambda})\equiv\det L(\lambda)$ is invariant along the $H_1$-flow,
we obtain the symmetry $F(\lambda)=\overline{F(\bar{\lambda})}$.
This symmetry property implies that the zeros of $F(\lambda)$ always appear in conjugate pairs along the $H_1$-flow.
We denote such pairs as $\left\{\beta_j,\bar{\beta}_j\right\}_1^N$.
Thus we may set
\begin{subequations}
\begin{eqnarray}
F(\lambda)= -\frac{1}{4}\frac{b(\lambda)}{\alpha(\lambda)}= -\frac{1}{4}\frac{R(\lambda)}{\alpha^2(\lambda)},\label{FL3a}
\\
L^{12}({\lambda})=-\left(\langle \mathbf{P},\mathbf{P}\rangle-\epsilon\overline{\langle \hat{\mathbf{Q}},\hat{\mathbf{Q}}\rangle}\right) \frac{m(\lambda)}{\alpha(\lambda)},\label{FL3b}
\\
L^{21}({\lambda})=\left(\langle \mathbf{Q},\mathbf{Q}\rangle-\epsilon\overline{\langle \hat{\mathbf{P}},\hat{\mathbf{P}}\rangle} \right)\frac{n(\lambda)}{\alpha(\lambda)},\label{FL3c}
\end{eqnarray}
\label{FL3}
\end{subequations}
where
\begin{subequations}
\begin{eqnarray}
\alpha(\lambda)= \prod_{j=1}^{N}(\lambda-\alpha_j)(\lambda-\bar{\alpha}_j),
~~b(\lambda)= \prod_{j=1}^{N}(\lambda-\beta_j)(\lambda-\bar{\beta}_j),\label{Rmn3a}
\\
m(\lambda)=\prod_{j=1}^{2N-1}( \lambda-\mu_j),~~n(\lambda)=\prod_{j=1}^{2N-1}( \lambda-\nu_j),\label{Rmn3b}
\\
R(\lambda)=\alpha(\lambda)b(\lambda)\equiv\prod_{j=1}^{2N}
(\lambda-\lambda_j)(\lambda-\bar{\lambda}_j),~~\lambda_j=\alpha_j,~~\lambda_{N+j}=\beta_j,~~j=1,\cdots,N. \label{Rmn3c}
\end{eqnarray}
\label{Rmn3}
\end{subequations}
The formula (\ref{Rmn3c}) implies the consideration of the following hyperelliptic curve $\Gamma$ of genus $g=2N-1$,
\begin{eqnarray}
\zeta^2 =4\prod_{j=1}^{2N}(\lambda-\lambda_j)(\lambda-\bar{\lambda}_j)
\equiv 4\prod_{j=1}^{4N}(\lambda-\lambda_j),~~\lambda_{2N+k}=\bar{\lambda}_k,~~1\leq k\leq 2N,
\label{HC3}
\end{eqnarray}
where $\lambda_j\neq \lambda_l$, $j\neq l$, $1\leq j, l\leq 4N$.
Let $a_1,\cdots,a_g$, $b_1,\cdots,b_g$ be canonical basis of cycles on $\Gamma$.
Denote by $A_{jk}$ the value of integral of the holomorphic differential $\tilde \omega_j$ along $a_k$, and denote by $\mathbf{C}_j$ the column vectors of the inverse matrix $\mathbf{C}=\left(\int_{a_k}\tilde{\omega}_j\right)^{-1}_{g\times g}$. As before, we define the vector of Abel-Jacobi variables $\mathbf{\Phi}$ and $\mathbf{\Psi}$ as the normalization of the quasi-Abel-Jacobi variables.

Inserting the symmetries $\alpha_{N+j}=\bar{\alpha}_j$, $\beta_{N+j}=\bar{\beta}_j$, $j=1,\cdots,N$, into (\ref{Omg}), (\ref{Lam}) and (\ref{hatS}), we obtain
\begin{subequations}
\begin{eqnarray}
\hat{S}_{k}=\sum_{j=1}^{2N}\left(\lambda_j^{k}+\bar{\lambda}_j^{k}\right),~~k\geq 1,
\label{hats3}
\\
\Lambda_0=1,~~\Lambda_{1}=\frac{1}{2}\hat{S}_1,
~~\Lambda_k=\frac{1}{2k}\left(\hat{S}_{k}+\sum_{i+j=k,~i,j\geq 1}\hat{S}_{i}\Lambda_j\right),
~~k\geq 2,
\label{Lam3}
\\
\boldsymbol{\Omega}_0=0,~~\boldsymbol{\Omega}_1=\frac{1}{2}\mathbf{C}_1,
~~\boldsymbol{\Omega}_k=\frac{1}{2}\sum_{i+j=k,~i\geq 0,~j\geq 1}\Lambda_{i}\mathbf{C}_{j},
~~k\geq 2.
\label{Omg3}
\end{eqnarray}
\label{hlo3}
\end{subequations}
By using the formulae (\ref{hlo3}) and the proposition \ref{proAJHXT}, we obtain
\begin{proposition}
In the Jacobi variety, the nonlocal Hamiltonian systems (\ref{HXnol}) and (\ref{HTnol}) are linearized by the Abel-Jacobi variables as
\begin{eqnarray}
\begin{split}
\frac{d \mathbf{\Phi}}{dx}=\frac{1}{2}\mathbf{C}_1, ~~\frac{d \mathbf{\Psi}}{dx}=-\frac{1}{2}\mathbf{C}_1,
\\
\frac{d \mathbf{\Phi}}{dt}=i\boldsymbol{\Omega}_{2}, ~~\frac{d \mathbf{\Psi}}{dt}=-i\boldsymbol{\Omega}_{2},
\end{split}
\label{H12s3}
\end{eqnarray}
where $\boldsymbol{\Omega}_{2}=\frac{1}{2}(\mathbf{C}_2+\frac{1}{2}\hat{S}_{1}\mathbf{C}_{1})$.
\end{proposition}

\begin{remark}
The hyperelliptic curve (\ref{HC3}), in comparison with (\ref{cure1}), admits an additional anti-holomorphic involution. This involution enables us to construct algebro-geometric solutions to the nonlocal NLS equation (\ref{nnls}) by virtue of our nonlocal integrable Hamiltonian systems. This issue will be discussed in detail in section 4.2.
\end{remark}

\section{Applications of the nonlocal Hamiltonian systems to the AKNS equation in the presence of the space-inverse reductions}

In this section, we study the construction of algebro-geometric solutions to the AKNS equation in the presence of the space-inverse reductions by virtue of our nonlocal Hamiltonian systems. As a consequence, we obtain algebro-geometric solutions to the AKNS equation with the Dirichlet and with the Neumann boundary conditions, and obtain algebro-geometric solutions to the nonlocal NLS equation.

\subsection{Algebro-geometric solutions to the AKNS equation with odd and even fields}

We now construct the algebro-geometric solutions to the AKNS equation that satisfy the space-inverse symmetry (\ref{oddevenuv}) by virtue of the finite-dimensional decomposition presented in section 3.1.
Recall that the hyperelliptic curve associated with the nonlocal Hamiltonian systems (\ref{HXD}) and (\ref{HTD}) is given by (\ref{HCD}).
This hyperelliptic curve admits the following holomorphic involution
\begin{equation}
\tau_h:~~(\lambda,\zeta)\rightarrow (-\lambda,\zeta).
\label{tauh}
\end{equation}
We select all the branch points $\lambda_j$ of the surface (\ref{HCD}) as real.
Without loss of generality, we assume
\begin{equation}
\lambda_1<\lambda_2<\cdots<\lambda_{g+1}<0,~~\lambda_{g+1+j}=-\lambda_{g+2-j}, ~~j=1,\cdots,g+1.
\end{equation}
We choose the canonical basis of cycles (see Fig. 1) such that it is transformed under the action of the involution (\ref{tauh}) by the rule (see e.g. \cite{BBEIM1994,SM2021,BI1989})
\begin{equation}
\tau_h \mathbf{a}=\mathcal{S}\mathbf{a},~~\tau_h \mathbf{b}=\mathcal{R}\mathbf{b},
\label{basrule}
\end{equation}
where $\mathbf{a}=(a_1,a_2,\cdots,a_g)^T$, $\mathbf{b}=(b_1,b_2,\cdots,b_g)^T$, and
\begin{equation}
\mathcal{R}=\mathcal{S}^T, ~~\mathcal{S}_{jk}=-\delta_{j,g+1-k},~~j,k=1,\cdots,g.
\label{SR}
\end{equation}

\begin{figure}
\centering
\includegraphics{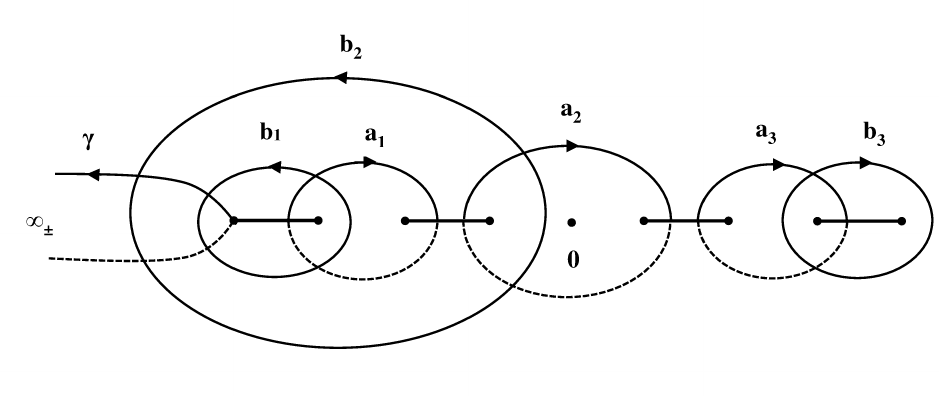}
\caption{\small{The chosen basis of cycles for the curve (\ref{HCD}) with $g=3$.
The parts of the cycles on the upper sheet are indicated by solid lines and the parts on the lower sheet are indicated by dashed lines.}}
\label{F1}
\end{figure}

The holomorphic involution (\ref{tauh}) and the transformation rule (\ref{basrule}) of the basis lead to the following symmetry relations:
\begin{subequations}
\begin{eqnarray}
\mathcal{S}^T\mathbf{C}=\mathbf{CJ},
\label{Csym}
\\
\mathbf{B}\mathcal{S}=\mathcal{R}\mathbf{B},
\label{BSRB}
\\
\mathcal{S}^T\mathbf{\Delta}=\mathbf{\Delta}+\mathbf{e},
\label{deltasym}
\\
N_1=I_1(\Gamma)=0,
\label{N1red1}
\end{eqnarray}
\label{CBdNsym}
\end{subequations}
where
\begin{equation}
\mathbf{J}=\diag((-1)^g,(-1)^{g-1},\cdots,-1),~~\mathbf{e}=\left(1,1,\cdots,1\right)^T.
\end{equation}
See Appendix A for the proof of the formulae (\ref{CBdNsym}).
Using (\ref{OmgD}) and (\ref{CBdNsym}) in (\ref{nlss1}), we obtain
\begin{eqnarray}
\begin{split}
u(x,t)=u(0,0)e^{N_2t}\frac{\theta\left(\frac{1}{2}\mathbf{C}_1x+\frac{i}{2}\mathbf{C}_2t+\mathbf{D}+\mathbf{\Delta}\right)
\theta\left(\mathbf{D}\right)}
{\theta\left(\frac{1}{2}\mathbf{C}_1x+\frac{i}{2}\mathbf{C}_2t+\mathbf{D}\right)\theta\left(\mathbf{D}+\mathbf{\Delta}\right)},
\\
v(x,t)=v(0,0)e^{-N_2t}\frac{\theta(\frac{1}{2}\mathbf{C}_1x+\frac{i}{2}\mathbf{C}_2t-\mathbf{D}-\mathbf{\Delta})
\theta(\mathbf{D})}
{\theta(\frac{1}{2}\mathbf{C}_1x+\frac{i}{2}\mathbf{C}_2t-\mathbf{D})\theta(\mathbf{D}+\mathbf{\Delta})},
\end{split}
\label{nlss1R}
\end{eqnarray}
where $\mathbf{C}_j$ satisfies the symmetry (\ref{Csym}) and $\mathbf{\Delta}$ satisfies the symmetry (\ref{deltasym}).

The last step is to choose a specialization of the vector of parameters $\mathbf{D}$ in the solution (\ref{nlss1R}) to ensure the oddness/evenness condition (\ref{oddevenuv}). We obtain the following result.
\begin{theorem}\label{prouvred}
Let $\Gamma$ be a hyperelliptic curve defined by (\ref{HCD}) with all the branch points $\lambda_j$ be real.
Let the canonical basis of $\Gamma$ be chosen such that it obeys the rule (\ref{basrule}) under the action of the involution (\ref{tauh}).
Choose the vector of parameters $\mathbf{D}$ such that it subjects to the restriction
\begin{equation}
\begin{split}
\mathbf{D}=\mathcal{S}^T\mathbf{D}+\mathbf{BM}+\mathbf{W},
\end{split}
\label{Dsym}
\end{equation}
where the vectors $\mathbf{M}\equiv (M_1,\cdots,M_g)^T\in \mathbb{Z}^g$ and $\mathbf{W}\equiv (W_1,\cdots,W_g)^T\in \mathbb{Z}^g$ satisfying $\mathcal{S}\mathbf{M}=-\mathbf{M}$, $\mathcal{S}^T\mathbf{W}=-\mathbf{W}$.
Then, for $M_{\frac{g+1}{2}}=2k$, $k\in \mathbb{Z}$, the algebraic-geometric solution (\ref{nlss1R}) satisfies the symmetry (\ref{oddevenuv}) with $\epsilon=1$. For $M_{\frac{g+1}{2}}=2k-1$, $k\in \mathbb{Z}$, the algebraic-geometric solution (\ref{nlss1R}) satisfies the symmetry (\ref{oddevenuv}) with $\epsilon=-1$.
\end{theorem}
{\bf Proof}
By using (\ref{Csym}), (\ref{deltasym}) and (\ref{Dsym}), we obtain
\begin{equation}
\begin{split}
-\frac{1}{2}\mathbf{C}_1x+\frac{i}{2}\mathbf{C}_2t+\mathbf{D}=\mathcal{S}^T\left(\frac{1}{2}\mathbf{C}_1x
+\frac{i}{2}\mathbf{C}_2t+\mathbf{D}\right) +\mathbf{BM}+\mathbf{W},
\\
-\frac{1}{2}\mathbf{C}_1x+\frac{i}{2}\mathbf{C}_2t+\mathbf{D}+\mathbf{\Delta}
=\mathcal{S}^T\left(\frac{1}{2}\mathbf{C}_1x+\frac{i}{2}\mathbf{C}_2t
+\mathbf{D}+\mathbf{\Delta}\right) +\mathbf{BM}+\mathbf{W}-\mathbf{e}.
\end{split}
\label{omg12sym}
\end{equation}
By using (\ref{omg12sym}) and the periodicity property of the Riemann theta function (see (\ref{thetapro})), we obtain
\begin{equation}
\varepsilon\equiv u(x,t)/u(-x,t)=\exp\left\{2\pi i\langle \mathbf{M}, \mathcal{S}^T\mathbf{\Delta}\rangle\right\},
\label{vare}
\end{equation}
where we have used the fact
$$\theta(\mathcal{S}^T\boldsymbol{\xi})=\theta(\boldsymbol{\xi})$$
which can be verified by virtue of (\ref{BSRB}).
By virtue of (\ref{deltasym}) and the symmetry $\mathcal{S}\mathbf{M}=-\mathbf{M}$, we find that the right hand side of (\ref{vare}) is equal to
\begin{equation}
\exp\left\{\pi iM_{\frac{g+1}{2}}\right\}.
\end{equation}
Thus, for $M_{\frac{g+1}{2}}=2k,~k\in \mathbb{Z}$, we obtain $\varepsilon=1$, which implies $u(x,t)=u(-x,t)$. For $M_{\frac{g+1}{2}}=2k-1,~k\in \mathbb{Z}$, we obtain $\varepsilon=-1$, which implies $u(x,t)=-u(-x,t)$. The argument for $v(x,t)$ can be performed via a very similar manner.  \QEDB

Based on theorem \ref{prouvred}, we immediately obtain the following assertion.
\begin{theorem}
Let the hyperelliptic curve and its canonical basis be defined as in theorem \ref{prouvred}.
Assume that the vector of parameters $\mathbf{D}$ subjects to the restrictions (\ref{Dsym}).
Then, for $M_{\frac{g+1}{2}}$ is even, the formulae (\ref{nlss1R}) solve the AKNS equation with the Neumann boundary condition (\ref{NBC}).
For $M_{\frac{g+1}{2}}$ is odd, the formulae (\ref{nlss1R}) solve the AKNS equation with the Dirichlet boundary condition (\ref{DBC}).
\end{theorem}

\begin{remark}
The Dirichlet and Neumann boundary conditions correspond to the limiting cases of Robin boundary conditions
\begin{equation}
\left.\left(u_x+cu\right)\right|_{x=0},~~c\in \mathbb{R},
\label{rbc}
\end{equation}
as $c\rightarrow \infty$ and as $c\rightarrow 0$, respectively. It is worth reminding that the algebraic-geometric solutions to the NLS equation with Robin boundary conditions (\ref{rbc}) for the cases $c\neq 0$ and $c\neq \infty$ were constructed in \cite{BI1989}. Here our results complete the cases $c\rightarrow 0$ and $c\rightarrow \infty$. The main difference with respect to the Baker function approach employed in \cite{BI1989} is that our approach is based on the finite-dimensional decomposition of the AKNS equation subjecting to space-inverse reductions.
\end{remark}

\begin{remark}
We required that the parameters $\alpha_j$ and thus the branch points of the corresponding hyperelliptic curve are all distinct. It has shown in \cite{BBEIM1994} that in the degeneration case (such as the parameters coincide in pairs) the algebraic-geometric solutions lead to the multi-soliton solutions. It will be interesting to study if the degeneracy of the algebraic-geometric solutions satisfying the Dirichlet/Neumann boundary conditions obtained in this subsection reproduces the previously known soliton solutions of the half-line problems with boundaries (see e.g. \cite{BH2009,BB2012,BT1991,Fokas,T1991,CZ2012,Zhang2019,CCD2022}). This topic is left for further study.
Here we only point out that in the degeneration situation (such as some of the parameters coincide) the finite-dimensional Hamiltonian systems considered in this paper involve multiple eigenvalue parameters and the superintegrability (see e.g. \cite{SI1,SI2,SI3,SI4,SI5,SI6}) will occur as a consequence.
\end{remark}

\subsection{Algebro-geometric solutions to the nonlocal NLS equation}

We now study the construction of algebro-geometric solutions to the nonlocal NLS equation (\ref{nnls}) by virtue of our nonlocal Hamiltonian systems (\ref{HXnol}) and (\ref{HTnol}). Recall that the associated hyperelliptic curve is given by (\ref{HC3}).
The starting point in this case is that (\ref{HC3}) admits the anti-holomorphic involution
\begin{equation}
\tau_a:~~(\lambda,\zeta)\rightarrow (\bar{\lambda},\bar{\zeta}).
\label{taua}
\end{equation}
We choose the canonical basis of cycles that is transformed under the action of the involution (\ref{taua}) by the rule
(see \cite{BBEIM1994,SM2021} for example)
\begin{equation}
\tau_a \mathbf{a}=-\mathbf{a},~~\tau_a \mathbf{b}=\mathbf{b}+\mathbf{\mathcal{T}}\mathbf{a},
\label{basrule2}
\end{equation}
where $\mathbf{a}=(a_1,a_2,\cdots,a_g)^T$, $\mathbf{b}=(b_1,b_2,\cdots,b_g)^T$ and
$\mathbf{\mathcal{T}}$ is the $g\times g$ matrix giving by
\begin{equation}
\mathbf{\mathcal{T}}\equiv (T_{jk})_{g\times g}=\left( \begin{array}{cccccc} 0 & 1 & 1 & \cdots & 1 & 1\\
1 & 0 & 1 &\cdots &1 &1 \\
\vdots & \vdots & \vdots &  & \vdots & \vdots \\
1 & 1 & 1 & \cdots & 0 & 1
\\
1& 1& 1 & \cdots & 1 & 0
\\ \end{array} \right).
\label{T}
\end{equation}
See Fig. 2 for the choice of the basis of the cycles.
The $a_j$ cycle circles around the branch cut between the pair $(\lambda_j,\lambda_{g+1+j})$ clockwise on the upper sheet.
The $b_j$ cycle starts on the upper sheet, moves clockwise through the branch cut between the pair $(\lambda_{g+1},\lambda_{2g+2})$, changing to the lower sheet, and returns to the upper sheet by passing through the $(\lambda_j,\lambda_{g+1+j})$ branch cut.

\begin{figure}
\centering
\includegraphics[width=0.8\linewidth]{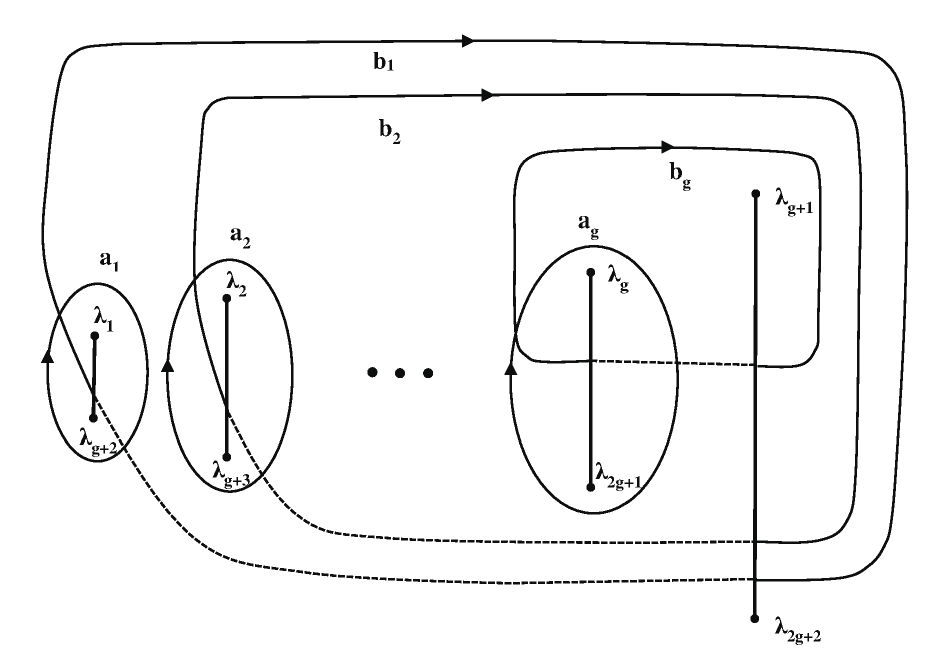}
\caption{\small{ The chosen basis of cycles for the curve (\ref{HC3}).
The parts of the cycles on the upper sheet are indicated by solid lines and the parts on the lower sheet are indicated by dashed lines.}}
\label{F2}
\end{figure}

The anti-holomorphic involution (\ref{taua}) and the transformation rule (\ref{basrule2}) lead to the following specialization of parameters of solutions (\ref{nlss1}):
\begin{subequations}
\begin{eqnarray}
\bar{\boldsymbol{\Omega}}_k=-\boldsymbol{\Omega}_k,~~k\geq 1,
\label{barsomg}
\\
\bar{\mathbf{B}}=-\mathbf{B}-\mathbf{\mathcal{T}}^T,
\label{barB}
\\
\bar{N}_1=N_1,
\label{barIN1}
\\
\bar{\mathbf{\Delta}}=- \mathbf{\Delta}+ \mathbf{e},
\label{bardelta}
\end{eqnarray}
\label{obndsym2}
\end{subequations}
where $\mathbf{\mathcal{T}}$ is defined by (\ref{T}) and $\mathbf{e}=\left(1,1,\cdots,1\right)^T$.
The proof for the above formulae will be presented in Appendix B.

By virtue of the above results, we obtain
\begin{theorem}\label{prouvnonred}
Let $\Gamma$ be a hyperelliptic curve defined by (\ref{HC3}).
Select the canonical basis of $\Gamma$ such that it obeys the rule (\ref{basrule2}) under the action of the involution (\ref{taua}).
Choose the initial values $u(0,0)$ and $v(0,0)$ and the vector of parameters $\mathbf{D}$ such that they subject to the restrictions
\begin{eqnarray}
v(0,0)=\epsilon\bar{u}(0,0),~~\epsilon=\pm 1,
\label{uv00}
\\
\bar{\mathbf{D}}=-\mathbf{D}+\mathbf{W},~~\mathbf{W}\in \mathbb{Z}^g.
\label{Dbar}
\end{eqnarray}
Let $u(x,t)$ and $v(x,t)$ be constructed in the form of the formulae (\ref{nlss1}) but in the context of the above chosen hyperelliptic curve and parameters.
Then such constructed algebraic-geometric solutions satisfy the symmetry
\begin{equation}
v(x,t)=\epsilon\bar{u}(-x,t).
\label{uvbar}
\end{equation}
In particular, such constructed $u(x,t)$ satisfies the nonlocal NLS equation (\ref{nnls}).
\end{theorem}
{\bf Proof}
We first prove that $(\ref{Dbar})$ and $(\ref{uv00})$ imply the important relation
\begin{equation}
\bar{N}_2=-N_2.\label{N2bar}
\end{equation}
To show this, we write $N_2$ as (recall that $N_2$ is defined by (\ref{DDelta}))
\begin{equation}
N_2=iN_{2}^{(1)}+2iN_{2}^{(2)}
+2iN_{2}^{(3)},
\end{equation}
where
\begin{equation}
N_{2}^{(1)}=\frac{1}{2}\hat{S}_2-I_2(\Gamma),
~~N_{2}^{(2)}=\sum_{j,k=1}^g\Omega_{1}^j\Omega_{1}^k\partial_{jk}^2\ln\theta(\mathbf{D}),
~~N_{2}^{(3)}=u(0,0)v(0,0).
\end{equation}
Using (\ref{barsomg}) and (\ref{barIN1}), we find $\bar{N}_{2}^{(1)}=N_{2}^{(1)}$.
Using (\ref{uv00}), we find $\bar{N}_{2}^{(3)}=N_{2}^{(3)}$.
Using (\ref{barsomg}), (\ref{Dbar}) and the periodicity property (\ref{thetapro}) of the $\theta$ function, we obtain
\begin{equation}
\begin{split}
\bar{N}_{2}^{(2)}-N_{2}^{(2)}=\sum_{j,k=1}^g\Omega_{1}^j\Omega_{1}^k\partial_{jk}^2\ln\frac{\theta(\mathbf{\bar{D}})}{\theta(\mathbf{D})}
=0.
\end{split}
\label{barN22}
\end{equation}
Thus $N_2$ does satisfy the identity (\ref{N2bar}).
The identity (\ref{barB}) implies
 \begin{eqnarray}
\overline{\theta(\boldsymbol{\xi})}=\theta(\bar{\boldsymbol{\xi}}), \quad \boldsymbol{\xi}\in\mathbb{C}^g.
\label{bartheta}
\end{eqnarray}
Using the identities (\ref{obndsym2}), (\ref{N2bar}) and (\ref{bartheta}),
we arrive at
\begin{eqnarray}
\bar{u}(-x,t)=\bar{u}(0,0)e^{-N_1x-N_2t}
\frac{\theta\left(\boldsymbol{\Omega}_1x+i\boldsymbol{\Omega}_2t+\bar{\mathbf{D}}-\mathbf{\Delta}\right)
\theta\left(\bar{\mathbf{D}}\right)}
{\theta\left(\boldsymbol{\Omega}_1x+i\boldsymbol{\Omega}_2t+\bar{\mathbf{D}}\right)
\theta\left(\bar{\mathbf{D}}-\mathbf{\Delta}\right)}.
\label{baru}
\end{eqnarray}
After straightforward calculations using (\ref{Dbar}), (\ref{baru}) and the periodicity property of the $\theta$ function,
we finally obtain
\begin{equation}
\frac{\bar{u}(-x,t)}{v(x,t)}=\frac{\bar{u}(0,0)}{v(0,0)}=\epsilon.
\label{uvbard}
\end{equation}
This completes the proof. \QEDB

The algebraic-geometric solutions discussed in this section has exponential growth with respect to $x$ as $N_1\neq 0$.
This exponential growth can be eliminated by using the holomorphic involution (\ref{tauh}) as considered in section 4.1. To show this, we select the branch points of the hyperelliptic surface (\ref{HC3}) further as
\begin{equation}
\lambda_j=-\lambda_{2g+3-j},~~1\leq j\leq \frac{g+1}{2}.
\label{lamsym}
\end{equation}
In this situation,  the canonical basis of cycles considered above is transformed under the action of the involution (\ref{tauh}) by the rule (see \cite{BBEIM1994,S1988,SM2021} for example)
\begin{equation}
\tau_h \mathbf{a}=\mathcal{S}\mathbf{a},~~\tau_h \mathbf{b}=\mathcal{Q}\mathbf{a}+\mathcal{R}\mathbf{b},
\label{basrule3}
\end{equation}
where the matrices $\mathcal{S}$, $\mathcal{Q}$ and $\mathcal{R}$ are given by
\begin{eqnarray}
\begin{split}
\mathcal{S}_{1k}=-1,~~\mathcal{S}_{jk}=\delta_{j,g+2-k},~~2\leq j\leq g,~~1\leq k\leq g,
\\
\mathcal{R}=\mathcal{S}^T, ~~\mathcal{Q}=\re(\mathbf{B})\mathcal{S}-\mathcal{R}\re(\mathbf{B}).
\label{SQR}
\end{split}
\end{eqnarray}
The transformation rule (\ref{basrule3}) leads to the following specialization of parameters (see Appendix C for the proof of these relations)
\begin{subequations}
\begin{eqnarray}
I_1(\Gamma)=0,
\label{I1hi}
\\
\mathcal{S}^T\boldsymbol{\Omega}_j=(-1)^j\boldsymbol{\Omega}_j, ~~1\leq j\leq g,
\label{Omghi}
\\
\mathcal{S}^T\mathbf{\Delta}=\mathbf{\Delta}+\left(\mathbf{B}-\mathbf{I}\right)\mathbf{e}_1,~~\mathbf{e}_1=\left(1,0,\cdots,0\right)^T.
\label{deltahi}
\end{eqnarray}
\label{iodsym3}
\end{subequations}
Using (\ref{I1hi}) and the identity $\hat{S}_{2k-1}=0$, $k\geq 1$, we immediately obtain $N_1=0$.
Thus, the exponential growth is eliminated.

\begin{remark}
The algebraic-geometric solution of the nonlocal NLS equation constructed by virtue of our nonlocal Hamiltonian systems coincides with the one found recently 
in \cite{SM2021} by using the Baker function approach. In other words, we provide an alternative way to construct algebraic-geometric solutions of integrable nonlocal soliton equations, which is based on the finite-dimensional decompositions of the corresponding nonlocal equations.
\end{remark}


\section{Conclusion and discussion}

We presented a procedure to implement the NLP method that permits to impose space-inverse reductions on soliton equations. As a result, we obtain new integrable nonlocal finite-dimensional Hamiltonian systems.
With aid of these nonlocal integrable Hamiltonian systems, we constructed algebraic-geometric solutions to the AKNS equation with Dirichlet/Neumann conditions, and constructed algebraic-geometric solutions to the nonlocal NLS equation.

We end this paper with the following remarks.
\begin{itemize}
\item [1)]
Let $\tau_k$ be the variable of the $\tilde{H}_k$-flow. It has been shown in \cite{CaoGengWu1999} that if $\left(\mathbf{p}(x,\tau_k),\mathbf{p}(x,\tau_k)\right)$ is a compatible solution of the $H_1$-flow and $\tilde{H}_k$-flow, then $u(x,\tau_k)=-\langle \mathbf{p}(x,\tau_k), \mathbf{p}(x,\tau_k) \rangle$, $v(x,\tau_k)=\langle \mathbf{q}(x,\tau_k), \mathbf{q}(x,\tau_k)\rangle$ solves the $n$-th equation in the AKNS hierarchy.
We note that we may consider the following reduction for the general $\tilde{H}_k$-flow,
\begin{eqnarray}
\begin{split}
\alpha_{N+j}=-\alpha_{j},~1\leq j\leq N,
\\
\left(p_{N+j}(x,\tau_k),q_{N+j}(x,\tau_k)\right)=\sqrt{\epsilon} \left(p_{j}(-x,(-1)^k\tau_k),-\epsilon q_{j}(-x,(-1)^k\tau_k)\right)
\equiv\sqrt{\epsilon} \left(\hat{p}_{j},-\epsilon\hat{q}_{j}\right).
\end{split}
\label{DRG}
\end{eqnarray}
We can check directly that the canonical equation of $\tilde{H}_k$-flow does admit the above reduction.
Indeed, by inserting (\ref{DRG}) into the polynomial integrals (\ref{Fk}) and (\ref{Hk}), we may deduce that
\begin{eqnarray}
\begin{split}
F_{k}(x,\tau_k)=(-1)^{k+1}F_{k}(-x,(-1)^k\tau_k),~\tilde{H}_{k}(x,\tau_k)=(-1)^{k+1}\tilde{H}_{k}(-x,(-1)^k\tau_k),~k\geq 0.
\end{split}
\label{FDks1}
\end{eqnarray}
By using the above symmetry relations, we may verify directly that the expressions for $\hat{q}_{j,\tau_k}, \hat{p}_{j,\tau_k}$ are consistent with those for  $q_{j,\tau_k}$, $p_{j,\tau_k}$. Using the above reduction we are possible to construct algebro-geometric solutions of higher order equations in the AKNS hierarchy with the potentials satisfying reverse-space/time symmetries or satisfying suitable boundary conditions. This issue will be investigated in the future.

\item [2)]
The integrable symplectic map is the discrete version of the finite-dimensional integrable Hamiltonian system (see e.g. \cite{M1991,BRST1991,R1991,CGW1999,Qiao1999,GDZ2007}).
The results presented in this paper can also be extended to the integrable symplectic map.
Precisely speaking, there also exist nonlocal (with respect to the discrete variable) counterparts for certain integrable symplectic maps, and the resulting nonlocal integrable symplectic maps can be used to construct algebro-geometric solutions to the nonlocal discrete soliton equations or to discrete soliton equations with certain boundary conditions. These results will be presented elsewhere.

\item [3)]
After this paper was finished, we learn that the nonlocal finite-dimensional system related to the nonlocal NLS equation (\ref{nnls}) with $\epsilon=-1$ was investigated recently in \cite{WD2023} by virtue of Lie-Poisson structure.
\end{itemize}


\section*{ACKNOWLEDGMENTS}

B. Xia was supported by the National Natural Science Foundation of China under Grant No. 12271221.
R. Zhou was supported by the National Natural Science Foundation of China under Grant No. 12171209.
Both the authors thank Dianlou Du for his helpful discussions.

\begin{appendices}

\section{The proof for the symmetry relations (\ref{CBdNsym})}
Let $\widehat{\boldsymbol{\omega}}=\tau_h\boldsymbol{\omega}$. Calculating $a$-period of $\widehat{\boldsymbol{\omega}}$, we obtain
\begin{equation}
\oint_{a_k}\widehat{\omega}_j=\oint_{\tau_h a_k}\omega_j=\sum_{l=1}^g \mathcal{S}_{kl}\oint_{a_l}\omega_j=\mathcal{S}_{kj}.
\label{hatomegaa1}
\end{equation}
On the other hand, using (\ref{omega}) we have
\begin{equation}
\begin{split}
\oint_{a_k}\widehat{\omega}_j=&\sum_{l=1}^g \oint_{a_k}\tau_h\left(C_{jl}\tilde{\omega}_l\right)
=\sum_{l=1}^g C_{jl}\oint_{a_k}\tau_h\left(\tilde{\omega}_l\right)
\\
=&\sum_{l=1}^g C_{jl}(-1)^{g+1-l}\oint_{a_k}\tilde{\omega}_l
=\sum_{l=1}^g C_{jl}(-1)^{g+1-l}A_{lk}=\left(\mathbf{CJA}\right)_{jk},
\end{split}
\label{hatomegaa2}
\end{equation}
where $\mathbf{J}=\diag((-1)^g,(-1)^{g-1},\cdots,-1)$.
Equations (\ref{hatomegaa1}) and (\ref{hatomegaa2}) imply that
\begin{equation}
\mathcal{S}^T=\mathbf{CJ}\mathbf{C}^{-1},
\label{SC}
\end{equation}
which is equivalent to (\ref{Csym}).
In addition, we have
\begin{equation}
\widehat{\boldsymbol{\omega}}=\tau_h\boldsymbol{\omega}=\mathbf{C}\tau_h\tilde{\boldsymbol{\omega}}
=\mathbf{C}\mathbf{J}\tilde{\boldsymbol{\omega}}
=\mathcal{S}^T\mathbf{C}\tilde{\boldsymbol{\omega}}
=\mathcal{S}^T\boldsymbol{\omega}.
\label{hatomrela}
\end{equation}
where the fourth equality uses (\ref{SC}).
Calculating $b$-period of $\widehat{\boldsymbol{\omega}}$, we obtain
\begin{equation}
\oint_{b_k}\widehat{\omega}_j=\oint_{\tau_h b_k}\omega_j=\sum_{l=1}^g \mathcal{R}_{kl}\oint_{b_l}\omega_j=(\mathcal{R}\mathbf{B})_{kj}.
\label{hatomegab1}
\end{equation}
On the other hand, using (\ref{hatomrela}) we have
\begin{equation}
\begin{split}
\oint_{b_k}\widehat{\omega}_j
=\sum_{l=1}^g \mathcal{S}_{lj}\oint_{b_k}\omega_l
=\sum_{l=1}^g \mathcal{S}_{lj}B_{lk}=\left(\mathbf{B}\mathcal{S}\right)_{kj}.
\end{split}
\label{hatomegab2}
\end{equation}
Equations (\ref{hatomegab1}) and (\ref{hatomegab2}) lead to (\ref{BSRB}).

To prove (\ref{deltasym}), we set
\begin{eqnarray}
\mathbf{\Delta}_{\pm}=\int_{P_{2g+2}}^{\infty_{\pm}}\boldsymbol{\omega},
\end{eqnarray}
where $P_{j}$, $1\leq j\leq 2g+2$, denote the branch points at $\lambda=\lambda_{j}$.
We note that $\tau_0\boldsymbol{\omega}=-\boldsymbol{\omega}$ implies that
\begin{eqnarray}
\int_{P_{B}}^{P}\boldsymbol{\omega}=-\int_{P_{B}}^{\tau_0P}\boldsymbol{\omega},
\label{tau0}
\end{eqnarray}
where $\tau_0$ is a hyperelliptic involution, $\tau_0: (\lambda,\zeta)\rightarrow (\lambda,-\zeta)$,
$P$ is a point on $\Gamma$, and $P_B$ denotes a branch point of $\Gamma$.
Using (\ref{tau0}), we obtain
\begin{eqnarray}
\mathbf{\Delta}=\mathbf{\Delta}_{+}-\mathbf{\Delta}_{-}=2\int_{P_{2g+2}}^{\infty_{+}}\boldsymbol{\omega},
\label{Delta2}
\end{eqnarray}
where the integration path is placed on the upper sheet of $\Gamma$.
Using (\ref{hatomrela}), (\ref{Delta2}) and $\tau_hP_{2g+2}=P_{1}$, we have
\begin{eqnarray}
\mathcal{S}^T\mathbf{\Delta}=2\int_{P_{2g+2}}^{\infty_{+}}\mathcal{S}^T\boldsymbol{\omega}
=2\int_{P_{2g+2}}^{\infty_{+}}\tau_h\boldsymbol{\omega}
=2\int_{P_{1}}^{\infty_{+}}\boldsymbol{\omega}
=2\int_{P_{1}}^{P_{2g+2}}\boldsymbol{\omega}+\mathbf{\Delta},
\label{SDelta}
\end{eqnarray}
where the third equality uses the identity $\tau_h\infty_{\pm}=\infty_{\pm}$ (this follows from the asymptotics $\zeta=\pm 2(\lambda^{g+1}+O(\lambda^g))$
in the vicinity of the infinite points $\infty_{\pm}$ and the fact that $g$ is odd).
Calculating the line integral $2\int_{P_{1}}^{P_{2g+2}}\boldsymbol{\omega}$, we obtain
\begin{eqnarray}
2\int_{P_{1}}^{P_{2g+2}}\boldsymbol{\omega}=
\sum_{j=1}^{g+1}\oint_{\alpha_j}\boldsymbol{\omega}+\sum_{j=1}^g\oint_{a_j}\boldsymbol{\omega}
=\sum_{j=1}^g\oint_{a_j}\boldsymbol{\omega}
=\mathbf{e}.
\label{p12g}
\end{eqnarray}
where $\alpha_j$, $1\leq j\leq g+1$, denote the cycles enclose the branch cuts $\left(\lambda_{2j-1},\lambda_{2j}\right)$ in clock-wise orientation,
and the second equality uses the fact $\sum_{j=1}^{g+1}\oint_{\alpha_j}\boldsymbol{\omega}=0$ (see e.g. \cite{BBEIM1994,FK2015}). Inserting (\ref{p12g}) into (\ref{SDelta}), we arrive at (\ref{deltasym}).

We introduce the notation
$\mathcal{I}_{jk}=\oint_{a_k}\lambda \omega_j,~~j,k=1,2,\cdots,g$.
The quantities $\mathcal{I}_{jk}$ satisfy the following symmetry
\begin{equation}
\sum_{l=1}^g\left(\mathcal{S}_{kl}\mathcal{I}_{jl}+\mathcal{S}_{lj}\mathcal{I}_{lk}\right)=0.
\label{symI}
\end{equation}
Indeed,
\begin{equation}
\begin{split}
\oint_{a_k}\tau_h \left(\lambda\omega_j\right)=\oint_{\tau_h a_k}\lambda\omega_j=\sum_{l=1}^g \mathcal{S}_{kl}\mathcal{I}_{jl},
\\
\oint_{a_k}\tau_h \left(\lambda\omega_j\right)=-\oint_{a_k}\lambda\tau_h \left(\omega_j\right)
=-\sum_{l=1}^g\mathcal{S}_{lj}\oint_{a_k}\lambda\omega_l
=-\sum_{l=1}^g\mathcal{S}_{lj}\mathcal{I}_{lk}.
\end{split}
\label{lamomg}
\end{equation}
Comparing the above two equations, we obtain (\ref{symI}).
Using (\ref{SR}) and (\ref{symI}), we obtain
\begin{equation}
I_1(\Gamma)=\sum_{j=1}^g\mathcal{I}_{jj}=0.
\label{I1gram}
\end{equation}
Using (\ref{I1gram}) and (\ref{hatsD}), we obtain (\ref{N1red1}).

\section{The proof for the symmetry relations (\ref{obndsym2})}
Computing the complex conjugate of the integrals of holomorphic differentials $\tilde \omega_j$ along $a$-cycles,
we have
\begin{equation}
\bar{A}_{jk}=\overline{\oint_{a_k}\tilde{\omega}_j}=\oint_{a_k}\bar{\tilde{\omega}}_j=\oint_{a_k}\tau_a\tilde{\omega}_j
=\oint_{\tau_aa_k}\tilde{\omega}_j=-\oint_{a_k}\tilde{\omega}_j=- A_{jk},
\label{barA}
\end{equation}
which implies that
\begin{equation}
\bar{\mathbf{A}}=- \mathbf{A},~~\bar{\mathbf{C}}=- \mathbf{C}.
\label{barAC}
\end{equation}
Formulae (\ref{hlo3}) and (\ref{barAC}) yield (\ref{barsomg}).
Using (\ref{barAC}) and the relations $\bar{\tilde{\boldsymbol{\omega}}}=\tau_a\tilde{\boldsymbol{\omega}}$ and $\boldsymbol{\omega}= \mathbf{C}\tilde{\boldsymbol{\omega}}$, we find
\begin{equation}
\bar{\boldsymbol{\omega}}=- \tau_a\boldsymbol{\omega}.
\label{barna}
\end{equation}
Using (\ref{barna}), we obtain
\begin{eqnarray}
\begin{split}
\bar{\mathbf{B}}_{k}&=\overline{\oint_{b_k} \boldsymbol{\omega}}=\oint_{b_k} \bar{\boldsymbol{\omega}}
=-\oint_{b_k} \tau_a\boldsymbol{\omega}=-\oint_{\tau_a b_k} \boldsymbol{\omega}
\\
&=-\oint_{b_k} \boldsymbol{\omega}-\sum_{j=1}^{g}T_{kj}\oint_{a_j} \boldsymbol{\omega}
=-\mathbf{B}_{k}-\left(\mathbf{\mathcal{T}}_k\right)^T,
\quad
k=1,\cdots,g,
\end{split}
\end{eqnarray}
where $\mathbf{\mathcal{T}}_k$ denotes the $k$-th row of the matrix $\mathbf{\mathcal{T}}$.
Thus relation (\ref{barB}) holds.
For the quantity $I_s(\Gamma)$, we have
\begin{eqnarray}
\begin{split}
\overline{I_s(\Gamma)}&=\sum_{j=1}^g \overline{\oint_{a_j}\lambda^s \omega_j}
=\sum_{j=1}^g \oint_{a_j}\bar{\lambda}^s \bar{\omega}_j
=-\sum_{j=1}^g \oint_{a_j}\bar{\lambda}^s \tau_a\omega_j
\\
&=-\sum_{j=1}^g \oint_{a_j}\tau_a\left(\lambda^s \omega_j\right)
=-\sum_{j=1}^g \oint_{\tau_a a_j}\lambda^s \omega_j
=\sum_{j=1}^g \oint_{ a_j}\lambda^s \omega_j=I_s(\Gamma).
\end{split}
\end{eqnarray}
Thus $\overline{I_s(\Gamma)}=I_s(\Gamma)$. Using this relation and $\bar{\hat{S}}_1=\hat{S}_1$, we find (\ref{barIN1}).

We now prove (\ref{bardelta}).
Using the relations $\tau_aP_{2g+2}=P_{g+1}$ and $\tau_a\infty_{\pm}=\infty_{\pm}$,
we have
\begin{subequations}
\begin{eqnarray}
\bar{\mathbf{\Delta}}_{+}=\int_{P_{2g+2}}^{\infty_{+}}\bar{\boldsymbol{\omega}}
=-\int_{P_{2g+2}}^{\infty_{+}}\tau_a\boldsymbol{\omega}
=-\int_{P_{g+1}}^{\infty_{+}}\boldsymbol{\omega}
=-\int_{P_{g+1}}^{P_{2g+2}}\boldsymbol{\omega}-\mathbf{\Delta}_{+},
\label{Deltapbar}
\\
\bar{\mathbf{\Delta}}_{-}=\int_{P_{2g+2}}^{\infty_{-}}\bar{\boldsymbol{\omega}}
=-\int_{P_{2g+2}}^{\infty_{-}}\tau_a\boldsymbol{\omega}
=-\int_{P_{g+1}}^{\infty_{-}}\boldsymbol{\omega}
=-\int_{P_{g+1}}^{P_{2g+2}}\boldsymbol{\omega}-\mathbf{\Delta}_{-},
\label{Deltambar}
\end{eqnarray}
\label{Deltabar}
\end{subequations}
where the paths of the line integrals in (\ref{Deltapbar}) and (\ref{Deltambar}) belong to the upper and the lower sheet of the surface $\Gamma$, respectively.
Using the fact that the sum of all $a_j$ cycles is homologous to the positively oriented contour around
the cut $(\lambda_{2g+2},\lambda_{g+1})$ (see for example \cite{BBEIM1994,FK2015}), we obtain
\begin{eqnarray}
2\int_{P_{2g+2}}^{P_{g+1}}\boldsymbol{\omega}=\sum_{j=1}^g\oint_{a_j}\boldsymbol{\omega}=\mathbf{e},
~~\mathbf{e}=\left(1,1,\cdots,1\right)^T,
\label{Pa}
\end{eqnarray}
where the line integral is taken to be on the upper sheet.
By using (\ref{Deltabar}) and (\ref{Pa}), we obtain
\begin{eqnarray}
\bar{\mathbf{\Delta}}=\bar{\mathbf{\Delta}}_{+}-\bar{\mathbf{\Delta}}_{-}=
-\left(\mathbf{\Delta}_{+}-\mathbf{\Delta}_{-}\right)+\mathbf{e}
=- \mathbf{\Delta}+\mathbf{e}.
\label{bardelta1}
\end{eqnarray}
Thus relation (\ref{bardelta}) holds.

\section{The proof for the symmetry relations (\ref{iodsym3})}
As in Appendix A we may show that the quantities $\mathcal{I}_{jk}$ satisfy the symmetry (\ref{symI})
but with the quantities $\mathcal{S}_{jk}$, $1\leq j,k\leq g$, replaced by (\ref{SQR}).
Setting $j=k=1$ in (\ref{symI}), we obtain
\begin{eqnarray}
2\mathcal{I}_{11}+\sum_{l=2}^g\mathcal{I}_{1l}=0.
\label{I11}
\end{eqnarray}
Setting $k=g+2-j$, $2\leq j\leq g$, in (\ref{symI}), we obtain
\begin{eqnarray}
\mathcal{I}_{jj}-\mathcal{I}_{1,g+2-j}+\mathcal{I}_{g+2-j,g+2-j}=0, ~~2\leq j\leq g.
\label{Ijj}
\end{eqnarray}
Adding equations (\ref{I11}) and (\ref{Ijj}) together, we find $\sum_{j=1}^g\mathcal{I}_{jj}=0$, and thus (\ref{I1hi}) holds.
To show (\ref{Omghi}), we first note that relations (\ref{lamsym}) imply
\begin{eqnarray}
\hat{S}_{2k-1}=0,~~\hat{S}_{2k}=2\sum_{j=1}^{N}\left(\lambda_j^{2k}+\bar{\lambda}_j^{2k}\right),~~k\geq 1,
\label{hatSh}
\end{eqnarray}
By calculating $a$-period of $\widehat{\boldsymbol{\omega}}\equiv \tau_h\boldsymbol{\omega}$ as performed in section 5.1, we obtain
\begin{eqnarray}
\mathcal{S}^T\mathbf{C}_j=(-1)^{g+1-j}\mathbf{C}_j,~~1\leq j\leq g,
\label{Chi}
\end{eqnarray}
where $\mathcal{S}$ is given by (\ref{SQR}).
Using (\ref{Omg3}), (\ref{hatSh}) and (\ref{Chi}), we obtain (\ref{Omghi}).
To derive (\ref{deltahi}), we use the fact (see for example \cite{FK2015,BK2011})
that the integrals of holomorphic differentials over $\mathbf{a}$ and $\mathbf{b}$ cycles can be expressed in terms of
line integrals between branch points
\begin{eqnarray}
\oint_{a_j}\boldsymbol{\omega}=2\int_{P_{g+j+1}}^{P_{j}}\boldsymbol{\omega},
~~
\oint_{b_j}\boldsymbol{\omega}=2\sum_{k=j}^{g-1}\int_{P_k}^{P_{g+k+2}}\boldsymbol{\omega}
+2\int_{P_{g}}^{P_{g+1}}\boldsymbol{\omega},
~~1\leq j\leq g,
\label{abline}
\end{eqnarray}
where all the line integrals are to be taken on the upper sheet of the surface $\Gamma$.
Proceeding as in section 5.1, we obtain
\begin{eqnarray}
\mathcal{S}^T\mathbf{\Delta}
=2\int_{P_{1}}^{P_{2g+2}}\boldsymbol{\omega}+\mathbf{\Delta}.
\label{SDelta2}
\end{eqnarray}
Using (\ref{abline}), we obtain
\begin{eqnarray}
\begin{split}
2\int_{P_{1}}^{P_{2g+2}}\boldsymbol{\omega}=&\oint_{b_1}\boldsymbol{\omega}+\sum_{j=2}^g\oint_{a_j}\boldsymbol{\omega}
+2\int_{P_{g+1}}^{P_{2g+2}}\boldsymbol{\omega}
\\
=&\oint_{b_1}\boldsymbol{\omega}+\sum_{j=2}^g\oint_{a_j}\boldsymbol{\omega}
-\sum_{j=1}^g\oint_{a_j}\boldsymbol{\omega}
\\
=&\oint_{b_1}\boldsymbol{\omega}-\oint_{a_1}\boldsymbol{\omega}
\\
=&\left(\mathbf{B}-\mathbf{I}\right)\mathbf{e}_1.
\end{split}
\label{P1g}
\end{eqnarray}
Inserting (\ref{P1g}) into (\ref{SDelta2}), we obtain (\ref{deltahi}).

\end{appendices}

\vspace{1cm}
\small{

}

\end{document}